\documentclass[12pt]{article}
\begin{document}
\textwidth=15.8cm
\textheight=21cm
\author{V. A. Fateev\thanks{On leave of absence from Landau 
Institute for Theoretical Physics, ul.
Kosygina 2, 117940 Moscow, Russia.}}
\title{Normalization Factors, Reflection Amplitudes and Integrable Systems}
\date{}
\maketitle

\begin{center}
Laboratoire de Physique Math\'{e}matique,

Universit\'{e} Montpellier II

Pl. E.Bataillon, 34095 Montpellier, France.
\end{center}

\bigskip\ 

\begin{center}
{\bf Abstract}
\end{center}

We calculate normalization factors and reflection amplitudes in the
W-invariant conformal quantum field theories. Using these CFT data we derive
vacuum expectation values of exponential fields in affine Toda theories and
related perturbed conformal field theories. We apply these results to evaluate 
explicitly the expectation values of order parameters in the field theories 
associated with statistical systems, like $XY$, $Z_n$-Ising and Ashkin-Teller 
models. The same results are used for the calculation of the 
asymptotics of cylindrically symmetric solutions of the classical Toda
equations which appear in topological field theories. 
The integrable boundary Toda theories are considered. We derive
boundary reflection amplitudes in non-affine case and boundary one point
functions in affine Toda theories. The boundary ground state energies are 
cojectured. In the last section we describe the duality 
properties and calculate the reflection amplitudes in 
integrable deformed Toda theories.

\section{ Introduction}

There is a large class of 2D quantum field theories (QFTs) which can be
considered as perturbed conformal field theories (CFTs). These theories are
completely defined if one specifies its CFT data and the relevant operator
which plays the role of perturbation. The CFT data contain explicit
information about the ultraviolet (UV) asymptotics of the field theory while
the long distance behavior is the subject of analysis. If a perturbed CFT
contains only massive particles it is equivalent to the relativistic
scattering theory and is completely defined specifying the S-matrix.
Contrary to CFT data the S-matrix data exhibit the information about the
long distance property of the theory in the explicit way while the UV
asymptotics have to be derived.

A link between these two kinds of data would provide a good view point for
the understanding of the general structure of 2D QFT. In general this
problem does not look tractable. Whereas the CFT data can be specified in a
relatively simple way the general S-matrix is rather complicated object even
in two dimensions. However, there exists rather important class of 2D QFTs
(integrable theories) where the scattering theory is factorized and 
S-matrix can be described in great details. These theories admit rather
complete description in the UV and IR regimes.

In this paper we consider the application of special 
CFT data (normalization factors and reflection amplitudes) to the 
analysis of integrable QFTs. The normalization factors appear in the
CFTs which possess the representation for the primary fields in terms of
vertex operators and are used for the proper normalization of the fields. We
show that rather important information about the integrable perturbed CFTs
is encoded at these data. As an example of integrable QFT we study the
simply-laced affine Toda theory (ATT), which can be considered as the
perturbed CFT (non-affine Toda theory). This CFT possesses the extended
symmetry generated by the $W$-algebra. We calculate the normalization
factors (NFs) and derive the reflection amplitudes for $W$-invariant CFTs.
We use the normalization factors to relate the coupling constants and the 
masses of particles in perurbed CFTs.
 
The reflection amplitudes (RAs) in CFT define the linear transformations
between the different exponential fields corresponding to the same primary
field of the full symmetry algebra of theory. They play the crucial role in the calculation of
the vacuum expectation values of the exponential fields in perturbed CFTs 
\cite{FLZ} as well as for the analysis of UV asymptotics of the 
observables in these QFTs \cite{ZZ},\cite{AFR}. 
In section 3 we use these reflection amplitudes to
calculate the one point functions in ATT and related perturbed CFTs. These 
results are applied to the explicit evaluation of the vacuum expectation values of the order parameters (spins) in critical $Z_n$ Ising and 
Ashkin-Teller models 
perturbed by the thermal operators. In section 4 we use the vacuum 
expectation values of the exponential fields in ATT    
to derive  the asymptotics of the cylindrically symmetric solutions of the
classical Toda equations which appear in the analysis of topological QFTs 
\cite{CV}. 

In section 5 we consider integrable boundary Toda theories. We derive
the boundary reflection amplitudes in the boundary non-affine Toda theories
(NATTs) and vacuum expectation values of boundary exponential fields in 
ATTs. Using these expectation values we calculate the classical boundary ground 
state energy and give the conjecture for it in the quantum case. The boundary 
scattering theory which is consistent with this conjecture is discussed. 
In the last section we describe the duality properties and derive the
reflection amplitudes for integrable deformed Toda theories, which have some 
physical applications.

\section{ Affine and Non-Affine Toda Theories, Normalization Factors and
Reflection Amplitudes}

The ATT corresponding to the Lie algebra $G$ of rank $r$ is described by the
action:

\begin{equation}
{\cal A}_b=\int     d^2x\left[ \frac 1{8\pi }(\partial _\mu \varphi )^2+\mu
\sum_{i=1}^re^{be_i\cdot \varphi }+\mu e^{be_0\cdot \varphi }\right] ,
\label{at}
\end{equation}
where $e_i,$ $i=1,...,r$ are the simple roots of Lie algebra $G$ and $-e_0$
is a maximal root:

\begin{equation}
e_0+\sum_{i=1}^rn_ie_i=0.  \label{mr}
\end{equation}
The fields $\varphi $ in eq. (\ref{at}) are normalized so that at $\mu =0$

\begin{equation}
\left\langle \varphi _a(x),\varphi _b(y)\right\rangle =-\delta _{ab}\log
\left| x-y\right| ^2  \label{n}
\end{equation}
We will consider later the simply-laced Lie algebras $A,D,E.$

For real $b$ the spectrum of these ATTs consists from $r$ particles with
masses $m_i\ (i=1,...,r)$ given by:

\begin{equation}
m_i=m\nu _i,\qquad m^2=\frac 1{2h}\sum_{i=1}^rm_i^2,  \label{s}
\end{equation}
where $h$ is Coxeter number of $G$ and $\nu _i^2$ are the eigenvalues of the
mass matrix:

\begin{equation}
M_{ab}=\sum_{i=1}^rn_i(e_i)^a(e_i)^b+(e_0)^a(e_0)^b.  \label{mab}
\end{equation}

The exact relation between the parameter $m$ characterizing the spectrum of
physical particles and the parameter $\mu $ in the action (\ref{at}) can be
obtained by the Bethe ansatz method (see for example \cite{ALZ},\cite{F}).
It can be easily derived from the results of paper \cite{F} and has the form:

\begin{equation}
-\pi \mu \gamma (1+b^2)=\left[ \frac{mk(G)\Gamma \left( \frac 1{(1+b^2)h}
\right) \Gamma \left(\frac{b^2}{(1+b^2)h}\right)b^2 }{2\Gamma \left( \frac 1h
\right)h(1+b^2) }\right] ^{2(1+b^2)},  \label{mmu}
\end{equation}
where as usual $\gamma (x)=\Gamma (x)/\Gamma (1-x)$ and

\begin{equation}
k(G)=\left( \prod_{i=1}^rn_i^{n_i}\right) ^{1/2h}  \label{k}
\end{equation}
with $n_i$ defined by the equation (\ref{mr}).

The ATTs can be considered as the perturbed CFTs. Without the last term
(with root $e_0$) the action (\ref{at}) describes NATTs, which are
conformal. To describe the generator of the conformal symmetry we introduce
the complex coordinates $z=x_1+ix_2,\ \bar{z}=x_1-ix_2$ and vector

\begin{equation}
Q=(b+1/b)\rho ,\qquad \rho =\frac 12\sum_{\alpha >0}\alpha ,  \label{q}
\end{equation}
where the sum in the definition of the Weyl vector $\rho $ runs over all
positive roots $\alpha $ of Lie algebra $G.$

The holomorphic stress energy tensor

\begin{equation}
T(z)=-\frac 12(\partial _z\varphi )^2+Q\cdot \partial _z^2\varphi  \label{se}
\end{equation}
ensures the local conformal invariance of the NATT with the central charge

\begin{equation}
c=r+12Q^2=r(1+h(h+1)(b+1/b)^2).  \label{cc}
\end{equation}

Besides the conformal invariance NATT possesses the additional symmetry
generated by two copies of the chiral $W(G)$-algebras: $W(G)\otimes $ $
\overline{W}(G)$. The full chiral $W(G)$-algebra contains $r$ holomorphic
fields $W_j(z)$ $(W_2(z)=T(z))$ with spins $j$ which follow the exponents of
Lie algebra $G.$ The explicit representation of these fields in terms of
fields $\partial _z\varphi _a$ can be found in \cite{FL}. The primary fields 
$\Phi _w$ of $W(G)$-algebra are specified by $r$ eigenvalues $w_j,\
j=1,...,r $ of the operators $W_{j,0}$ (the zeroth Fourier components of the
currents $W_j(z)$):

\begin{equation}
W_{j,0}\Phi _w=w_j\Phi _w;\qquad W_{j,n}\Phi _w=0,\quad n>0.  \label{w}
\end{equation}
The exponential fields:

\begin{equation}
V_a(x)=e^{a\cdot \varphi (x)}  \label{ex}
\end{equation}
are spinless conformal primary fields with the dimensions:

\begin{equation}
\Delta (a)=w_2(a)=\frac{Q^2}2-\frac{(a-Q)^2}{2}.  \label{dim}
\end{equation}
The fields $V_a(x)$ are also primary fields with respect to all chiral
algebra $W(G)$ with the eigenvalues $w_j$ depending on $a$. The functions 
$w_j(a)$ possess the symmetry with respect to Weyl group ${\cal W}$ of Lie
algebra $G$ \cite{FL} which acts at the vector $a$ as:

\begin{equation}
a\rightarrow s(a)=Q+{\bf \hat{s}}(a-Q){\bf \ ,\quad \hat{s}\in }{\cal W}
;\qquad w_j(s(a))=w_j(a).  \label{wg}
\end{equation}
It means that the fields $V_{s(a)}$ for different ${\bf\hat{s}}\in {\cal W}$ 
are the reflection image of each other and are related by the linear 
transformation:

\begin{equation}
V_a(x)=R_s(a)V_{s(a)}(x),  \label{ra}
\end{equation}
where $R_s(a)$ is the ''reflection amplitude''.

To calculate the function $R_s(a)$ we introduce the normalized primary
fields $\Phi _w:$

\begin{equation}
\Phi _w=N^{-1}(a)V_a(x),  \label{no}
\end{equation}
where the normalization factor $N(a)$ is chosen in the way that field $\Phi
_w$ satisfies conformal normalization condition:

\begin{equation}
\left\langle \Phi _w\left( x\right) ,\Phi _w\left( y\right) \right\rangle =
\frac 1{\left| x-y\right| ^{4\Delta }}.  \label{cnc}
\end{equation}
The normalized fields $\Phi _w$ are invariant under the reflection
transformations and hence:

\begin{equation}
R_s(a)=\frac{N(a)}{N(s(a))}.  \label{rnr}
\end{equation}

For the calculation of the normalization factor $N(a)$ we can use the
integral representation for the correlation functions of the $W(G)$
-invariant CFT (see \cite{FL} for details). We note that the operators $
\hat{Q}_i$ defined as:

\begin{equation}
\hat{Q}_i=\int d^2xe^{be_i\cdot \varphi (x)}  \label{sc}
\end{equation}
commute with all of the elements of $W(G)$-algebra and can be used for the
calculation of the correlation functions in NATT. In particular, if vector 
$ a $ satisfies the condition 
\begin{equation}
2a+\sum_{i=1}^rk_ie_i=0  \label{con}
\end{equation}
with non-negative integer $k_i$ we obtain from eqs.(\ref{no}, \ref{cnc}) the
following expression for the function $N(a)$ in terms of Coulomb integrals 
\cite{FL}:

\begin{equation}
N^2(a)=\left| x\right| ^{4\Delta }\left\langle V_a(x)V_a(0)\prod_{i=1}^r
\frac{\hat{Q}_i^{k_i}}{k_i!}\right\rangle  \label{int}
\end{equation}
where the expectation value in eq.(\ref{int}) is taken over the Fock vacuum
of massless fields $\varphi $ with the correlation functions (\ref{n}).

The normalization integral (\ref{int}) can be calculated and the result has
the form:

\begin{eqnarray}
N^2(a)&=&(\pi \mu \gamma (b^2))^{-2\rho \cdot a/b}
\nonumber \\
&&\times
\prod_{\alpha >0}\frac{
\Gamma (1+Q_\alpha /b)\Gamma (1+Q_\alpha b)\Gamma (1+\bar{a}_\alpha
/b)\Gamma (1+\bar{a}_\alpha b)}{\Gamma (1-Q_\alpha /b)\Gamma (1-Q_\alpha
b)\Gamma (1-\bar{a}_\alpha /b)\Gamma (1-\bar{a}_\alpha b)}  \label{nf}
\end{eqnarray}
where:

\begin{equation}
\bar{a}=(a-Q),\quad Q_\alpha =Q\cdot \alpha ,\quad \bar{a}_\alpha
=(a-Q)\cdot \alpha  \label{abar}
\end{equation}
and the product runs over all positive roots $\alpha $ of Lie algebra 
$G.$

With these normalization factors the integral representation with screening
charges \cite{FL} reproduces correlation functions 
of the primary fields $\Phi _w$ (\ref{no})
satisfying the conformal normalization condition (\ref{cnc}). We accept 
eq.(\ref{nf}) as the proper analytical continuation of the function $N^2(a)$ 
for all $a$. It gives the following expression for the reflection amplitude 
$R_s(a):$

\begin{equation}
R_s(a)=\frac{N(a)}{N(s(a))}=\frac{A_{s(a)}}{A_a},  \label{rsa}
\end{equation}
where 
\begin{equation}
A(a)=(\pi \mu \gamma (b^2))^{\rho \cdot \bar{a}/b}\prod_{\alpha >0}\Gamma (1-
\bar{a}_\alpha /b)\Gamma (1-\bar{a}_\alpha b).  \label{amp}
\end{equation}
With this function the reflection relation (\ref{ra}) can be written in more
symmetric form as:

\begin{equation}
A_aV_a(x)=A_{s(a)}V_{s(a)}(x),\qquad s(a)=Q+{\bf \hat{s}}(a-Q),\quad {\bf 
\hat{s}}\in {\cal W}.  \label{rere}
\end{equation}

The action ${\cal A}_\beta $ of the imaginary ATT can be obtained from the
action (\ref{at}) by the substitution $b\rightarrow i\beta ,\quad \mu
\rightarrow -\mu .\,$ This action is invariant under the transformation 
$\varphi \rightarrow \varphi +2\pi \theta /\beta ,$ where $\theta $ belongs
to the weight lattice of $G.$ It means that the space of the vacua of
imaginary ATT is equivalent to this lattice. In the classical case this
theory possesses the solitons. The basic solitons have the topological charge
proportional to the weights of the fundamental representations $\pi _i(G).$
In the quantum case the particles, corresponding to these solitons form the
multiplets which can be associated with the fundamental representations of
the Yangian $Y_i(G)$. The masses $M_i$ $(i=1,...,r)$ of the particles in
these multiplets have the form: $M_i=M\nu _i,$ where mass parameter $M$ and
eigenvalues $\nu _i$ are defined by the eqs.(\ref{s},\ref{mab}). The exact
relation between the parameters of the action ${\cal A}_\beta $ and the
physical mass $M$ can be obtained by the Bethe ansatz method \cite{ALZ},\cite
{F} and has the form:

\begin{equation}
\pi \mu \gamma (1-\beta ^2)=\left[ \frac{M\pi k(G)\Gamma \left( \frac 1{
(1-\beta ^2)h}\right) }{\Gamma \left( \frac 1h\right) \Gamma \left( \frac{
\beta ^2}{(1-\beta ^2)h}\right) }\right] ^{2(1-\beta ^2)}.  \label{immu}
\end{equation}

The imaginary ATTs are non-unitarian QFTs. However, for special vealues of 
 $\beta ^2=p/(p+1)$
with integer \ $p>h,$ these QFTs admit the restriction with respect to the
affine quantum group $U_q(\hat{G})$ with level equal to zero and 
$q=\exp(i\pi /\beta ^2)$ to the unitarian QFTs \cite{RS},\cite{BL}. 
This restricted theories can be 
considered as the minimal unitarian models 
${\cal M}_p(G)$ of the $W(G)$- invariant
CFTs perturbed by the relevant operator $\Phi _{ad}\in {\cal M}_p(G)$
associated with the adjoint representation of $G$. 

The minimal models ${\cal M}_p(G)$ are described in details in \cite{FL}. 
They are characterized by the central charges:

\begin{equation}
c_p(G)=r\left( 1-\frac{h(h+1)}{p(p+1)}\right) ,\quad p>h.  \label{cch}
\end{equation}
The primary fields $\Phi (\Omega ,\Omega ^{\prime })\in {\cal M}_p(G)$ are
specified by two highest weight vectors $\Omega $ and $\Omega ^{\prime }$
which satisfy the conditions: $-\Omega \cdot e_0\leq p+1-h$ and  
$-\Omega ^{\prime}\cdot e_0\leq p-h$. 
The fields $\Phi _\Omega =\Phi (\Omega ,0)$ and $\Phi
_{\Omega ^{\prime }}=\Phi (0,\Omega ^{\prime })$ form two closed operator
subalgebras of ${\cal M}_p(G)$ CFTs. The primary, strongly degenerate 
fields $\Phi (\Omega,\Omega ^{\prime })$, normalized by the condition 
(\ref{cnc}) can be
represented in terms of the fields $\varphi $ of imaginary NATT by the
relation:

\begin{equation}
\Phi (\Omega ,\Omega ^{\prime })={\cal N}^{-1}(a)e^{ia\cdot \varphi },\quad
a=-\beta \Omega +\Omega ^{\prime }/\beta ,\quad \beta ^2=p/(p+1).
\label{repr}
\end{equation}
The normalization factors ${\cal N}(a)$ can be obtained from the NFs $N(a)$  
(\ref{nf}) by the substitutions:

\begin{equation}
b\rightarrow i\beta ,\quad \ a\rightarrow ia,\quad \ \mu \rightarrow -\mu
,\quad Q\rightarrow iQ_\beta =i(\beta -1/\beta )\rho .  \label{sbs}
\end{equation}
Being rather complicated for general values of $a,$ the NFs ${\cal N}(a)$
and $N(a)$ simplify for some particular values of $a$, which are useful for
practical applications. For example, it happens if $\ a=-\beta \omega _i$,
where $\omega _i$ are the fundamental weights of $G\ (\omega _i\cdot
e_j=\delta _{ij})$ satisfying the condition: $-e_0\cdot \omega _i=1$ (or $
n_i $ in the eq.(\ref{mr}) is equal to $1$). In this case the 
normalization factors have the
form which is very similar to Weyl formula (modulo $\gamma $-functions) for
the inverse dimensions of representations $\pi _i(G)$:

\begin{equation}
{\cal N}^2(-\beta \omega _i)=(\pi \mu \gamma (u))^{2\omega _i\rho
}\prod_{\alpha >0}\frac{\gamma (\rho _\alpha u)}{\gamma ((\rho +\omega
_i)_\alpha u)};\quad u=1-\beta ^2.  \label{gd}
\end{equation}
In particular, the NFs ${\cal N}(-\beta \omega _i)$ for Lie algebra $A_{n-1}$
(where all fundamental weights satisfy this condition) can be written as:

\begin{equation}
{\cal N}^2(-\beta \omega _k)=(\pi \mu \gamma (u))^{k(n-k)}\prod_{j=1}^k\frac{
\gamma (ju)}{\gamma ((n+1-j)u)},  \label{an}
\end{equation}
For the perturbing operator $\Phi _{ad}\equiv \Phi _{\Omega _{ad}}=\Phi
_{-e_0}$ the normalization factor for all simply-laced cases can be written
as:

\begin{displaymath}
{\cal N}^2(\beta e_0)=(\pi \mu \gamma (u))^{2h-2}\frac{(1-(h+1)u)^2\gamma
(u)\gamma (qu)\gamma (\frac{h-2q+2}2u)}{(1-hu)^2\gamma (\frac{h+2q-2}{2}u)
\gamma ((h-q)u)\gamma ((h+1)u)}  
\end{displaymath}
here $u=1-\beta ^2=1/(p+1)$ and $q=\max_in_i,$ where $n_i$ are defined by
the eq.(\ref{mr}): $q(A)=1,\ q(D)=2,\ q(E_6)=3,\ q(E_7)=4,\ q(E_8)=6.$

The action of the perturbed CFT ${\cal M}_p(G)$ can be written in the form:
\begin{equation}
{\cal A}_p={\cal A}_{CFT}+\lambda \int d^2x\Phi _{ad}(x).  \label{pcft}
\end{equation}
The perturbing operator normalized by eq.(\ref{cnc}) can be 
represented using eq.(\ref{repr}) as: 
$\Phi _{ad}={\cal N}^{-1}(\beta e_0)\exp (i\beta e_0\cdot \varphi )$. 
This exponential field is
invariant under the quantum group restriction of ATT \cite{RS}. Comparing
the action (\ref{pcft}) with the action of the imaginary ATT we can write 
$\lambda =-\mu {\cal N}(\beta e_0).$ Expressing $\mu $ in terms of the
physical mass parameter $M$ by eq.(\ref{immu}) we obtain the exact relation
between coupling constant $\lambda $ and $M$ \cite{F}:
\begin{eqnarray}
 \lambda ^2 &=&
\left[ \frac{M\pi k(G)\Gamma \left( \frac{p+1}h\right) }{\Gamma 
\left( \frac1h\right)\Gamma \left( \frac ph\right) }\right] ^{4hu} 
\nonumber \\
&& \times
\frac{(1-hu)^{-2}(1-(h+1)u)^2\gamma (qu)\gamma (\frac{h-2q+2}2u)}
{\pi^2\gamma (u)\gamma (\frac{h+2q-2}2u)\gamma ((h-q)u)\gamma ((h+1)u)}. 
\label{lm}
\end{eqnarray}
This relation permits us to express the UV perturbative expansions in
coupling constant $\lambda $ in terms of the IR parameter $M.$

\section{Expectation Values of Local Fields in ATT and Related Perturbed CFT}

The vacuum expectation values (VEVs) of local fields play an important role
in the QFT and statistical mechanics. In statistical mechanics the VEVs
determine the ''generalized susceptibilities'', i.e. the linear response of
the system to external fields. In the QFT defined as a perturbed CFT the
VEVs provide all the information about its correlation functions that is not
accessible through direct calculation in conformal perturbation theory \cite
{ALZM}. Recently, some progress was made in the calculation of the VEVs in
two dimensional integrable QFTs \cite{LZ},\cite{FLZ}. Namely, it was shown
in \cite{FLZ} that VEVs of the exponential fields $V_a(x)$ in perturbed CFTs
satisfy the same ''reflection relations'' as the vertex operators $V_a(x)$
in basic CFT.

We define the function $G(a)$ as the vacuum expectation value of the
operator $\exp (a\cdot \varphi )$ in ATTs with real coupling $b.$

\begin{equation}
G(a)=\left\langle \exp (a\cdot \varphi )\right\rangle _b .  \label{g}
\end{equation}
For any element ${\bf \hat{s}}$ of Weyl group ${\cal W}$ this function
satisfies the functional equation:

\begin{equation}
A_aG(a)=A_{s(a)}G(s(a)),\quad s(a)=Q+{\bf \hat{s}}(a-Q),\quad {\bf \hat{s}
\in }{\cal W}  \label{fe}
\end{equation}
where function $A_a$ is given by eq.(\ref{amp}).

The minimal meromorphic solution to these functional equations 
which respects all
the symmetries of extended Dynkin diagrams of Lie algebras $ADE$ has the
form \cite{VFA}:

\begin{eqnarray}
G(a) & =  & 
   \left[ \frac{mk(G)\Gamma \left( \frac 1{(1+b^2)h}\right) 
\Gamma \left(\frac{b^2}{(1+b^2)h}\right)b^2 }
{2\Gamma \left( \frac 1h\right)h(1+b^2) }\right]
^{-a^2}   \nonumber \\
  & & \times\exp \left( \int\limits_0^\infty \frac{dt}t[a^2e^{-2t}-F_b(a,t)]
\right)  \label{vev}
\end{eqnarray}
where:

\begin{eqnarray}
F_b(a,t)&=&\sinh ((1+b^2)t)
\nonumber \\
&&\times
\sum_{\alpha >0}\frac{\sinh (ba_\alpha t)\sinh
((b(a-2Q)_\alpha +h(1+b^2))t)}{\sinh t\sinh (b^2t)\sinh ((1+b^2)ht)}.
\label{fb}
\end{eqnarray}
This solution satisfies many possible perturbative and non-perturbative
tests for one point function in ATTs. For example, it can be easily derived 
from eq.(\ref{mmu}) and equation of motion that bulk vacuum energy $E(b)$ in 
ATTs is expressed in terms of function $G(a)$ as:

\begin{equation}
n_i E(b)=h(1+b^2)\mu G(be_i).  \label{ebulk}
\end{equation}
The values of function $G(a)$ for $a$ equal to $be_i$ can be calculated 
explicitly and the result coincides with the known expression for the bulk 
vacuum energy \cite{DdV}:

\begin{equation}
E(b)=\frac{m^2 \sin(\pi/h)}{4\sin(\pi x/h)\sin(\pi (1-x)/h)};  \qquad 
x=\frac{b^2}{1+b^2}.  \label{ebul}
\end{equation}
Some other nonperturbative tests for one poin functions will be considered 
bellow. 

The VEVs of the exponential fields in imaginary ATTs:

\begin{equation}
{\cal G}(a)=\left\langle \exp (ia\cdot \varphi )\right\rangle _\beta
\label{gim}
\end{equation}
can be obtained from function $G(a)$ by the the substitution (\ref{sbs}). Being expressed in terms of
physical mass parameter $M$, it has a form:

\begin{equation}
{\cal G}(a)=\left[ \frac{M\pi k(G)\Gamma \left( \frac 1{(1-\beta ^2)h}
\right) }{\Gamma \left( \frac 1h\right) \Gamma \left( \frac{\beta ^2}
{(1-\beta ^2)h}\right) }\right] ^{a^2}\exp \left( -\int\limits_0^\infty 
\frac{dt}t[a^2e^{-2t}-
{\cal F}_\beta (a,t)]\right)  \label{ivev}
\end{equation}
with

\begin{equation}
{\cal F}_\beta (a,t)=\sinh u\sum_{\alpha >0}\frac{\sinh (\beta a_\alpha
t)\sinh ((\beta (a-2Q_\beta )_\alpha -hu)t)}{\sinh t\sinh (\beta ^2t)\sinh
(uht)}  \label{ifb}
\end{equation}
where $u=1-\beta ^2$ and $Q_\beta =(\beta -1/\beta )\rho .$

Being rather complicated for general $a$ functions ${\cal F}_\beta (a,t)$
and $F_b(a,t)$ simplify drastically for special directions of vector $a$,
which are useful for the practical applications. Namely, for $a=v\omega _k,$
where $\omega _k$ is a fundamental weight of $G,$ satisfying the condition: 
$-e_0\cdot \omega _k=1,$ function ${\cal F}_\beta (a,t)$ can be written as:

\begin{equation}
{\cal F}_\beta (v\omega _k,t)=\frac{\sinh ^2(\beta vt)}{\sinh t\sinh (\beta
^2t)}\left( 2\cosh (tu)\delta _{ij}-{\bf I}_{ij}\right) _{kk}^{-1}.
\label{inc}
\end{equation}
where the matrix ${\bf I}_{ij}=2\delta _{ij}-e_i\cdot e_j.$ In particular,
for Lie algebra $A_{n-1}$ all fundamental weights satisfy this condition and
we have:

\begin{equation}
{\cal F}_\beta (v\omega _k,t)=\frac{\sinh ^2(\beta vt)\sinh (kut)\sinh
((n-k)ut)}{\sinh t\sinh (\beta ^2t)\sinh (ut)\sinh (nut)}.  \label{fan}
\end{equation}

As an example of the application of eq.(\ref{ivev}), we consider the
particular correlations in $A_{n-1}$ imaginary ATT in the limit 
$n\rightarrow \infty .$ In this limit this QFT can be interpreted as the
special case of 3D $U(1)$ or $XY$ model. The action of this QFT can be
written in terms of the fields $U_k(x)=\exp (i\beta \varphi _k(x))\in U(1),\
k=1,...n,$ which satisfy periodic boundary condition: $U_1(x)=U_n(x).$
Namely:

\begin{equation}
{\cal A}_\beta =\int d^2x\left( \sum_{k=1}^n\frac 1{8\pi \beta ^2}\partial
_\mu U_k\partial _\mu U_k^{-1}+\mu U_kU_{k+1}^{-1}\right).  \label{xy}
\end{equation}
The model is continuous in two dimensions and discrete in the third. Using
eq.(\ref{ivev}) we can calculate the correlations between the fields 
$U_k(x)$ taken at the same point $x$ and at different $k.$ At the limit 
$n\rightarrow \infty $ we obtain that for $l>0$:

\begin{equation}
\left\langle U_{k\pm l}(x)U_k^{-1}(x)\right\rangle =
{M\beta ^2 \overwithdelims() 2}
^{2\beta ^2}\Gamma ^2(u)\frac{\Gamma (2+(l-1)u)\Gamma (1+(l-1)u)}{\Gamma
(1+lu)\Gamma (2+(l-2)u)}.  \label{cxy}
\end{equation}

The vacuum expectation values (\ref{ivev}) can be 
used for the calculations of the VEVs of the
primary operators in perturbed CFTs ${\cal M}_p(G)$ (\ref{pcft}). Here we
consider only the case of the subalgebra $\Phi _\Omega$. The primary fields
from this subalgebra are represented by the exponential fields (\ref{repr})
(with $\Omega^\prime$=0)
which are invariant under the quantum group restriction \cite{FLZ} and their
VEVs can be easily expressed through the function ${\cal G}(a)$ and NF 
${\cal N}(a)$. The general case we suppose to consider in the separate
publication.

The QFT (\ref{pcft}) contains the finite set $\{s\}$ of the degenerate
vacuum states, which can be specified by the highest weights $\theta _s$
satisfying the condition:

\begin{equation} 
-e_0\cdot \theta _s\leq p-h.  \label{iNql}
\end{equation}
The particles in
this QFT are the kinks interpolating between different vacua. The masses of
these excitations coincide with the masses of basic particles in imaginary
ATT and are related with coupling constant $\lambda $ by eq.(\ref{lm}). The
vector $\theta _s$ determines the shift of the field $\varphi :\ \varphi
\rightarrow \varphi +2\pi \theta _s/\beta $. corresponding to vacuum state $s$. 
As a result we obtain from the eq.(\ref{repr}) the following expression
for the VEVs of the normalized fields $\Phi _\Omega $ in the QFT (\ref{pcft}):

\begin{equation}
\left\langle \Phi _\Omega \right\rangle _s=\exp (-i2\pi \theta _s\cdot
\Omega ){\cal N}^{-1}(-\beta \Omega ){\cal G}(-\beta \Omega ).  \label{wv}
\end{equation}

As an example, we apply this equation to 
calculate the vacuum expectation values of
the spin field (order parameter) $\sigma $ in the critical $Z_n$-Ising models 
perturbed by the first thermal operator 
$\varepsilon$. $Z_n$-Ising models are the natural generalizations of Ising model to the case when spin variable takes its values 
in group $Z_n$ (see ref. \cite{VFAZ} for details). They are 
self-dual, i.e. possess Kramers-Wannier symmetry. At the self-dual manifold of
parameters $Z_n$-Ising models have the critical points, which were found in
\cite{VFAZ}. At these critical points   
$Z_n$-Ising models are described by the 
$Z_n$-parafermionic CFTs with central charge 
$c=\frac{2(n-1)}{n+2}$.
Besides the parafermionic symmetry these CFTs possess also $W(A_{n-1})$
symmetry and can be described by ${\cal M}_{n+1}(A_{n-1})$ minimal models
with $\beta ^2=\frac{n+1}{n+2}$ ($u=\frac 1{n+2}$). The spin field $\sigma $
in critical $Z_n$- Ising models has the dimension 
$\Delta =\frac{(n-1)}{2n(n+2)}$ and coincides with the primary field 
$\Phi _{\omega _1}\in {\cal M}_{n+1}(A_{n-1})$. 
The perturbing operator $\Phi _{ad}$ is exactly the
first thermal operator $\varepsilon $ with dimension 
$\delta =\frac 2{(n+2)}$. The operator $\varepsilon $ is anti self-dual, i.e. it changes the sign under duality transformation. It means that depending 
on the sign of $\lambda$ in eq.(\ref{pcft}) the perturbed theory will be in 
ordered or disordered phase. In the first (second) case the order (disorder) 
parameter $\sigma$ ($\mu$) has non-zero VEVs.

In the ordered phase the vacua $\{s\}$ 
are specified by the highest
weights $\theta _s$ which satisfy the inequality (\ref{iNql}), 
which can be written as:
 $-e_0\cdot \theta _s\leq1$. 
This condition has $n$ solutions $\theta _s=\{0,\omega _1,...,\omega
_{n-1}\}$, so, we have for the first factor in eq.(\ref{wv}): 
$\exp (-2\pi i\theta _s\cdot \omega _1)=
\exp (i2\pi s/n),\ s=0,...n-1.$ Taking ${\cal N}^{-1}(-\beta \omega _1)$ and 
${\cal G}(-\beta \omega _1)$ from eqs.(\ref{an},\ref{fan}) with $u=1-\beta^2=
\frac 1{n+2}$ 
we obtain the final
expression for VEV of the order parameter $\sigma $:
\begin{eqnarray}
\left\langle \sigma \right\rangle _s &=
&\exp(i2\pi s/n)\left[ \frac{M\pi \Gamma
\left( \frac{n+2}n\right) }{\Gamma \left( \frac 1n\right) \Gamma \left( 
\frac{n+1}n\right) }\right] ^{2\Delta }
{\gamma (\frac n{n+2}) \overwithdelims() \gamma (\frac 1{n+2})}
^{\frac 12}
\nonumber \\
&&\times \exp \left( \int\limits_0^\infty \frac{dt}t
\left( \frac{\sinh (\beta^2t)
\sinh ((n-1)ut)}
{\sinh t\sinh (nut)}-\frac{n-1}{n}\beta^2e^{-2t}\right) \right)  
\label{sszn}
\end{eqnarray}
For odd $n=2l+1$ the
integral in eq (\ref{sszn}) can be calculated and result has a form:

\begin{displaymath}
\left\langle \sigma \right\rangle _s=e^{i2\pi s/n}\left[ \frac{M\pi \Gamma
\left( \frac{n+2}n\right) }{\Gamma \left( \frac 1n\right) \Gamma \left( 
\frac{n+1}n\right) }\right] ^{2\Delta }{n+2 \overwithdelims() n}
^{\frac{n-1}{2n}}{\gamma (\frac n{n+2}) \overwithdelims() \gamma (\frac 1{n+2})}
^{\frac 12}\prod_{j=1}^l\frac{\gamma (\frac{2j-1}{n+2})}{\gamma (\frac{2j+1}{n+2})}.
\end{displaymath}
We note that particles in this theory can be considered as the solitons
connecting different vacua. In particular, their masses are proportional to
distances between these vacua: $M_k\sim \left| \exp (2\pi i(s+k)/n)-\exp
(2\pi is/n)\right| =2\sin (\pi k/n)$.

As another example of the application of the eq.(\ref{wv}) we consider the
critical Ashkin Teller (AT) model perturbed by the thermal operator. The
spin variable $\sigma $ ($\sigma ^{+}$) in this lattice model takes the
values in the group $Z_4$, which determines the symmetry of AT model. This
model has the critical line (in two dimensional space of parameters), where
it can be described by CFT. This critical line can be parametrized by the
conformal dimension $\Delta _\varepsilon =\gamma ^2$ of the thermal operator 
$\varepsilon $. The central charge of the corresponding CFT is equal to one
along the whole critical line. This CFT can be described by the action of a
massless free scalar field $\phi $ (See ref. \cite{ZAT} for details). The
thermal operator $\varepsilon $ can be represented through this 
field as $\varepsilon =\sqrt{2}\cos (\gamma \phi )$. 
The action of the AT model
perturbed by this operator has the form of Sine-Gordon QFT:

\begin{equation}
\mathcal{A}_{SG}=\int d^2x\left( \frac{(\partial _\mu \phi )^2}{16\pi }
+\lambda \sqrt{2}\cos (\gamma \phi ).\right)   \label{asG}
\end{equation}

The order parameters in the critical AT model are the spin fields 
$\sigma $ ($\sigma ^{+}$) and $\Sigma \sim \sigma ^2$. 
The field $\sigma $ ($\sigma ^{+}$) has conformal dimension 
$\Delta _\sigma =1/16$ which is independent on 
$\gamma $ along the whole critical line. 
This field is non-local with respect
to the field $\phi $ which has square-root branch point at the position of 
$\sigma $. The field $\Sigma $ is local with respect to $\phi $. It has
conformal dimension $\Delta _\Sigma =\gamma ^2/4$ and can be represented in
terms of $\phi $ as $\Sigma =\pm \sqrt{2}\cos (\gamma \phi /2)$.

At the points $\gamma ^2=1/n$ the critical AT model can be described by 
$\mathcal{M}_{h+1}(D_n)$ minimal models 
with $\beta ^2=\frac{2n-1}{2n}$ ($u=\frac 1{2n}$) 
which  have the central charge $c=1$ for all $n$. The
perturbing operator $\Phi _{ad}$ $\in \mathcal{M}_{h+1}(D_n)$ has the
conformal dimension  $\Delta _\varepsilon =1/n=\gamma ^2$ and is exactly the
thermal operator $\varepsilon $. The operator $\sigma $ ($\sigma ^{+}$) can
be represented by the field $\Phi _{\omega _n}(\Phi _{\omega _{n-1}})\in 
\mathcal{M}_{h+1}(D_n)$ for odd $n$ and by the field $\exp (i\frac \pi
4)(\Phi _{\omega _n}+i\Phi _{\omega _{n-1}})/\sqrt{2}$ for even $n$. Here 
$\omega _n,\omega _{n-1}$ denote the fundamental weights of the two spinor
representations of $D_n$. For all $n$ these fields have the conformal
dimension $\Delta _{\sigma}=1/16$. The operator $\Sigma $ 
with dimension $\gamma^2/4$ can be written as 
$\Sigma =\Phi _{\omega _1}$, where $\omega _1$ is
the fundamental weight of the vector representation.

In the ordered phase the vacua $\{s\}$ in the perturbed CFT (\ref{pcft}) are
specified by the highest weights $\theta _s$ which satisfy the inequality
(\ref{iNql}). For perturbed  $\mathcal{M}_{h+1}(D_n)$ CFTs this condition has
four solutions, namely, $\theta _s=\{0,\omega _1,\omega _{n-1},\omega _n\}$ in
agreement with $Z_4$ symmetry of AT model. The expectation values of the
order parameters $\sigma $ and $\Sigma $ can be calculated now, using
eq.(\ref{wv}). For this calculation it is convenient to use eqs.(\ref{gd}) and
(\ref{inc}), which for Lie algebra $D_n$ can be written as:

\begin{eqnarray}
\mathcal{N}^2(-\beta \omega _{n-1}) &=&\mathcal{N}^2(-\beta \omega
_n)=\left( \pi \gamma (u)\right) ^{n(n-1)/2}\prod\limits_{i=1}^{[n/2]}
\frac{\gamma ((2i-1)u)}{\gamma ((2n-2i)u)};  \nonumber \\
\mathcal{N}^2(-\beta \omega _1) &=&\left( \pi \gamma (u)\right) ^{2(n-1)}
\frac{\gamma (u)\gamma ((n-1)u)}{\gamma (nu)\gamma (2(n-1)u)};  \label{nDnn}
\end{eqnarray}
and

\begin{eqnarray}
\mathcal{F}_\beta (-\beta \omega _{n-1}) &=&\mathcal{F}_\beta (-\beta \omega
_n)=\frac{\sinh (\beta ^2t)\sinh (nut)\sinh (hut/2)}{\sinh t\sinh (2ut)\sinh
(hut)}  \nonumber \\
\mathcal{F}_\beta (-\beta \omega _n) &=&\frac{\sinh (\beta ^2t)\cosh
((n-2)ut)}{t\sinh t\cosh (hut/2)}.  \label{fDnn}
\end{eqnarray}
After simple transformations the vacuum expectation values for the order
parameters can be expressed through the soliton mass $M$ in the Sine-Gordon
theory and parameter $\gamma ^2=2u$. Namely, we obtain:

\begin{eqnarray}
\left\langle \sigma \right\rangle _s &=&\exp (i\pi s/2)\left( \frac{2M
\sqrt{\pi }\Gamma \left( \frac 1{2(1-\gamma ^2)}\right) }
{\Gamma \left( \frac{\gamma ^2}{2(1-\gamma ^2)}\right) }\right) ^{1/8}  
\nonumber \\
&&\times \exp \left( \int\limits_0^\infty \frac{dt}{8t}
\left( \frac{\cosh (2(1-\gamma ^2)t)}
{\cosh t\cosh \gamma ^2t\cosh ((1-\gamma ^2)t)}-e^{-2t}\right) \right) .
\label{sAT}
\end{eqnarray}
In this form eq.(\ref{sAT}) can be generalized to the arbitrary values of
parameter $\gamma ^2<1.$ In particular, at $\gamma ^2=2/3$ the expectation
value of field $\sigma $ was calculated in \cite{PBVF}, using different
approach. The result is in exact agreement with eq.(\ref{sAT}).

The expectation values of the operator $\Sigma =\Phi _{\omega _1}$,
calculated from eq.(\ref{wv}) have the form:

\begin{eqnarray}
\left\langle \Sigma \right\rangle _s &=
&(-1)^s\sqrt{2}\left( \frac{M\sqrt{\pi }\Gamma 
\left( \frac 1{2(1-\gamma ^2)}\right) }{2\Gamma 
\left( \frac{\gamma ^2}{2(1-\gamma ^2)}\right) }\right) ^{\gamma ^2/2}  
\nonumber \\
&&\times \exp \left( \int\limits_0^\infty \frac{dt}t
\left( \frac{\sinh (\gamma ^2t)}{2\sinh
t\cosh ((1-\gamma ^2)t)}-\frac{\gamma ^2}2e^{-2t}\right) \right)   \nonumber
\\
&=&\pm \sqrt{2}\left\langle \cos (\gamma \phi /2)\right\rangle _{SG}
\label{ssAT}
\end{eqnarray}
The last equality shows that this VEV coincides with the
expectation value for the field  $\pm\sqrt{2}\cos (\gamma \phi )$ 
in Sine-Gordon model calculated in \cite{LZ}. 
It gives us nonperturbative test for the
eq.(\ref{wv}).

At the end of this section we note that there are other nonperturbative
tests for eq.(\ref{wv}). For example, the CFTs $\mathcal{M}_{h+1}(E_{6,7})$
perturbed by operator $\Phi _{ad}$ coincide with tricritical $Z_3$ Potts and
tricritical Ising models perturbed by the first thermal operator. For the
case of $G=E_8$ this theory coincide with Ising model in the magnetic field.
The same models can be described as the minimal models of CFT, i.e. 
$\mathcal{M}_p(A_1)$ $(p=6,4,3)$ perturbed by the operator $\Phi _{12}$. 
The expectation values of the fields in the minimal models perturbed by the
operator $\Phi _{12}$ where calculated in \cite{FLZ}, using the quantum
restriction of imaginary Bullough-Dodd model. The calculation of the
expectation values of the fields in perturbed CFTs $\mathcal{M}_{h+1}(E)$
gives the exact agreement with the results of paper \cite{FLZ}.

\section{ Asymptotics of Cylindrical Solutions of Classical Toda Equations}

In this section we consider the application of VEVs $G(a)$ of the
exponential fields to the analysis of the special class of the solutions of
classical Toda equations, which appears in the topological QFTs \cite{CV}.
These solutions possess the cylindrical symmetry (i.e. depend only on 
$r=\left| x\right| $):

\begin{equation}
\partial _r^2\phi +r^{-1}\partial _r\phi =m^2\sum_{i=0}^rn_ie_i\exp
(e_i\cdot \phi );\qquad n_0=1,  \label{clt}
\end{equation}
and satisfy the following asymptotic conditions:

\begin{equation}
\phi \rightarrow -2a\log (mr)+B(a)\qquad at~\ r\rightarrow 0;  \label{sda}
\end{equation}

\begin{equation}
\phi \rightarrow \sum_{i=1}^r\eta _iX_i(a)K_0(\nu _imr)\qquad at\ \
r\rightarrow \infty ,  \label{lda}
\end{equation}
where $K_0(t)$ is the Mc-Donald function and $\eta _i$ are the eigenvectors
of mass matrix (\ref{mab}): $M\eta _i=\nu _i^2\eta _i,$ satisfying the
normalization condition:

\begin{equation}
\eta _{\bar{i}}=\bar{\eta}_i;\quad \bar{\eta}_i\cdot \eta _i=h;\quad 
\mathop{\rm Re}
\eta _i\cdot \rho >0.  \label{nrml}
\end{equation}
The solutions with these properties exist, if vector $a$ satisfies the
conditions: $a\cdot e_i<1,\ i=0,...r.$ In this case all other terms in the
short distance expansion (\ref{sda}) can be determined from the first two
using eq.(\ref{clt}). The constant term $B(a)$ in eq.(\ref{sda}) is not
arbitrary. It is defined by the global properties of eq.(\ref{clt}). Only
for special values of vector $B$ the solution has no singularities and
decrease exponentially at infinity. To determine the function $B(a)$ we
consider the semiclassical limit ($b\rightarrow 0$) of the QFT (\ref{at}).
At this limit we have (see eq.(\ref{mmu})) that $\mu =(mk(G))^2/4\pi
b^2+O(1) $ and after rescaling of field $\varphi :\ b\varphi =\tilde{\varphi}
$ the action (\ref{at}) is proportional to $1/b^2$:
\begin{equation}
{\cal A}_b=\frac{1}{4\pi b^2} \int d^2x
\left[ \frac 1{2 }(\partial _\mu\tilde{\varphi} )^2+(mk(G))^2
\sum_{i=0}^re^{e_i\cdot\tilde{ \varphi}}\right]+O(1).
\label{aTTT}
\end{equation} 
The classical equations
with cylindrical symmetry following from this action 
coincide with eqs.(\ref{clt}) after the shift:
$\phi =\tilde{\varphi}-\tilde{\varphi}_0 $, where
constant term $\tilde{\varphi}_0$ corresponds to the classical vacuum of ATT
(\ref{at}). It can be written in terms of the fundamental weights $\omega _i$
as:

\begin{equation}
\tilde{\varphi}_0=b\varphi _0=\sum_{i=1}^r(\log n_i-2\log k(G))\omega _i
\label{clv}.
\end{equation}

It is easy to see now that solution $\phi $ to eq.(\ref{clt}) 
can be expressed through the
semiclassical limit of the following two point function (which is completely 
determined by the saddle point contribution):

\begin{equation}
\phi +\tilde{\varphi}_0=\lim_{b\rightarrow 0} 
{\left\langle b\varphi (x)\exp (a\cdot \varphi (0)/b)\right\rangle _b \overwithdelims() G(a/b)}.
\label{cls}
\end{equation}
The asymptotics (\ref{sda}) is governed by the exponential term in the
correlation function and asymptotics (\ref{lda}) follows from form-factor
expansion for (\ref{cls}). The vector $B(a)$ can be now derived from the
operator product expansion:

\begin{equation}
\frac{\left\langle b\varphi (x)e^{a\cdot \varphi (0)/b}\right\rangle _b}
{G(a/b)}=-2a\log r+\frac{\left\langle b\varphi (0)e^{a\cdot \varphi
(0)/b}\right\rangle _b}{G(a/b)}+O(r^\sigma ).  \label{ope}
\end{equation}
It means that:

\begin{equation}
\phi +\tilde{\varphi}_0=-2a\log r+\lim_{b\rightarrow 0}b^2\partial _a\log
G(a/b)+O(r^\sigma );\quad \sigma >0.  \label{bb}
\end{equation}
This limit can be calculated explicitly with the result:

\begin{equation}
\phi +\tilde{\varphi}_0=-2a\log \left( \frac{mrk(G)}{2h}\right)
-\sum_{\alpha >0}\alpha \log \gamma \left( \frac{(\rho -a)_\alpha }h\right)
+O(r^\sigma ).  \label{rs}
\end{equation}
At $a=0$ the solution $\phi $ vanishes, so we have two different expressions
for $\tilde{\varphi}_0,$ namely, eq.(\ref{rs}) at $a=0$ and eq.(\ref{clv}).
Comparing them we obtain the amusing relation for gamma-functions:

\begin{equation}
\prod_{\alpha >0}\left( \gamma 
{\rho _\alpha  \overwithdelims() h}
\right) ^{-\alpha \cdot e_j}=n_j\left( k(G)\right) ^{-2},  \label{gfr}
\end{equation}
which is valid for all simply laced Lie algebras. As the final expression
for the constant term in eq.(\ref{sda}) we have:

\begin{equation}
B(a)=-2a\log {k(G) \overwithdelims() 2h}
-\sum_{\alpha >0}\alpha \left( \log \gamma \left( \frac{(\rho -a)_\alpha }h
\right) -\log \gamma \left( \frac{\rho _\alpha }h\right) \right).  \label{baf}
\end{equation}
The same expression for the asymptotics of $A_n-$ Toda solutions was
obtained by completely different method in ref.\cite{TW}.

The long distance asymptotics (\ref{lda}) can be derived from the one
particle form-factors in ATT (see ref.\cite{SL} for details). At the
semiclassical limit $b\rightarrow 0$ these form-factors are related with the
coefficients $X_i(a)$ as:

\begin{equation}
\left\langle 0\left| \exp (a\cdot \varphi /b)\right| A_i\right\rangle =G(a/b)
{h \overwithdelims() 2\pi }
^{1/2}X_i(a)\left( 1+O(b^2)\right).  \label{ff}
\end{equation}

It can be derived from form-factor equations that functions $X_i(a)$ are
equal to the characters of the fundamental representations $Y_i(G)$ of the
Yangian $Y(G):$

\begin{equation}
X_i(a)=Tr_{Y_i}\exp \left( \frac{2\pi i}h(a-\rho )\cdot H\right).  \label{xi}
\end{equation}
The functions $X_i(a)$ can be expressed through the similar characters $\chi
_i(a)$ taken over the fundamental representations $\pi _i(G)$ of Lie algebra 
$G.$ For all $i$ that satisfy the condition: $-e_0\cdot \omega _i=1$
functions $X_i(a)$ coincide with characters $\chi _i(a)$. For $A_n$ this is
valid for all $i.$ For $D_n$ we have that: $X_1=\chi _1,X_{n-1}=\chi
_{n-1},X_n=\chi _n,$ and for other representations:

\begin{equation}
X_{2j}(a)=1+\sum_{s=1}^j\chi _{2s}(a);\quad X_{2j+1}(a)=\sum_{s=0}^j\chi
_{2s+1}(a).  \label{dn}
\end{equation}
For the fundamental representations of $D$ and $E$ that coincide with
adjoint the functions $X_{ad}(a)$ are equal to $1+\chi _{ad}(a).$ The list
of expressions for functions $X_i$ in terms of the characters $\chi _i$ can
be found in \cite{K}. It can be shown that functions 
$X_i(a)$ possess the properties:

\begin{equation}
X_i(0)=0,\quad \frac h{2\pi }\partial _aX_i(0)=\eta _i;\quad X_i
{\rho  \overwithdelims() h+1}=1,  \label{pro}
\end{equation}
where $\eta _i$ are the eigenvectors of mass matrix (\ref{mab}) satisfying
normalization conditions (\ref{nrml}).

At the special points $a=\rho /(h+1)$ all $X_i$ are equal to one. At these
points the short distance expansion for the solution $\phi $ also simplifies
drastically (all higher terms in the asymptotics (\ref{sda}) are given by
the regular series in $(mr)^{2/(h+1)}$). These points appear in connection
with special class of topological QFT. In particular, the solutions of 
eq.(\ref{clt}) for these values of $a$ describe "the new sypersymmetric index" 
\cite{CV} for the imaginary ATT at $N=2$ sypersymmetric points: $\beta
^2=h/(h+1)$ \cite{FLMW}. We note that ATT give us an example of the system 
where representations of Yangian appear already in the classical case. 
We suppose to return to the analysis of eq. 
(\ref{clt}) in the separate publication.

\section{Boundary Toda Theories, Reflection Amplitudes,
One Point Functions and Ground State Energies}

At the previous sections we considered Toda theories defined in the whole
plane $R^2$. Here we consider simply-laced NATT and ATT 
defined at the half-plane 
$H=(x,y;y>0)$ with the integrable boundary conditions.

The integrability
conditions for the classical simply-laced ATT on $H$ were studied in the
paper\cite{BCD}. It was shown there that the action of integrable ATT can be
written as:

\begin{eqnarray}
{\cal A}_{bound}&=&\int\limits_Hd^2x\left[ \frac 1{8\pi }(\partial _\mu
\varphi )^2+\mu \sum_{i=1}^re^{be_i\cdot \varphi }+\mu e^{be_0\cdot \varphi
}\right]
\nonumber \\
&&
 +\mu _B\int dx\sum\limits_0^rd_ie^{be_i\cdot\varphi /2}  \label{atb}
\end{eqnarray}
where or all the parameters $d_i=0$, that corresponds to the Neumann
boundary conditions:

\begin{equation}
\partial _y\varphi (x,0)=0;  \label{N}
\end{equation}
or the parameters $d_i=\pm 1$ and the parameter $\mu _B$ is related with the
parameter $\mu $ in the bulk (in classical case) as:

\begin{equation}
\mu _B^2=\mu /\pi b^2.  \label{ND}
\end{equation}
For Lie algebra $A_1$ (Sinh-Gordon model) the integrability conditions are
much less restrictive and parameters $d_0$ and $d_1$ can have arbitrary
values \cite{GZ}. The background CFT for this case is the boundary Liouville
theory. The reflection amplitudes in boundary Liouville CFT for arbitrary
values of parameter $\mu _B$ or ($d_1$) and VEVs in boundary 
Sinh-Gordon model for
arbitrary values of $d_1$ and $d_0$ were found in \cite{FZZA} .
Here we discuss this problem for other Lie algebras where the choice of
integrable conditions is rather restrictive. We consider the Toda theories
with Neumann boundary conditions and in the case when all parameters $d_i=1$
(with $d_0=0$ in non-affine case). Really, these two quite different
classical theories in the quantum case are described by the same
theory and are related by duality transformation ($b\rightarrow 1/b$)
\cite{CRG}, \cite{GHB}. The
cases corresponding to different signs of parameters $d_i$ are more subtle
and will be considered elsewhere.

We start from the consideration of the boundary NATTs which are described by
the action (\ref{atb}) without the last term in the bulk action and $d_0=0$
in the boundary term. The boundary ATTs will be considered as perturbed
boundary CFTs.  At the whole plane NATTs possess the
infinite symmetry generated by two copies of chiral $W(G)$-algebras. These 
$ W(G)$-algebras contain $r$ holomorphic and $r$ antyholomorphic currents 
$W_j(z)$ and $\overline{W}_j(\overline{z})$ with spins that follow the
exponents of Lie algebra $G.$ At the half-plane with $W$-invariant boundary
conditions we have only one $W$-algebra. In this case the currents 
$\overline{W}_j(\overline{z})$ should be the analytical continuations of the
currents $W_j(z)$ to the lower half-plane. In particular, they should
coincide at the boundary. These conditions impose very strong restrictions
to the form of the boundary terms in the action. It is rather easy to derive
from the explicit form of $W$-currents \cite{FL} that Neumann boundary
conditions (\ref{N}) preserve $W$-symmetry. The boundary condition (\ref{ND}) 
whose quantum modified version has a form \cite{FZZA}:

\begin{equation}
\mu _B^2=\frac \mu 2\cot \left( \frac{\pi b^2}2\right)
\label{MND}
\end{equation}
with all $d_i=1$ and $d_0=0$ describes the dual theory and, hence, 
also preserve $W-$symmetry. 

In the
boundary Liouville CFT \cite{FZZA} the equation (\ref{MND}) was obtained as
the condition that boundary exponential fields corresponding to the
degenerate representations of Virasoro algebra satisfy the null vectors
equations. It can be shown that the same condition (\ref{MND}) is valid for
boundary exponential fields in NATT corresponding to the strongly degenerate
representations of the $W$-algebra. The null vectors equations simplify
drastically the OPEs of exponential primary fields with strongly degenerate 
fields. In these cases only the finite number of primary fields appear in OPEs 
and corresponding structure constants can be expressed in terms of the 
integrals with the screening charges. We will use this
property for the calculation of the boundary reflection amplitudes. We note
that in boundary Liouville theory the condition that null vectors equations 
for degenerate boundary fields are fulfilled is not necessary for conformal 
invariance and here it is, 
probably, necessary for preserving of $W$-symmetry.

In the $W$-invariant boundary NATT we have 
two kinds of the exponential fields. The
bulk fields $V_a(x,y)$ and the boundary fields $B_a(x)$ defined as:

\begin{equation}
V_a(x,y)=\exp \{a\cdot \varphi (x,y)\};\quad B_a(x)=\exp \{a\cdot \varphi
(x)/2\}.  \label{VA}
\end{equation}
These fields are specified by the same $r$ eigenvalues $w_j(a)$ that and
corresponding fields (\ref{ex}) defined on the whole plane. In particular,
their dimensions are given by eq.(\ref{dim}). The functions $w_j(a)$ are
invariant under the action of the Weyl group of $G$ (see section 2), defined
by eq.(\ref{wg}) and, hence, we can introduce the boundary reflection
amplitudes ${\cal R}_s(a)$ as:

\begin{equation}
B_a(x)={\cal R}_s(a)B_{s(a)}(x).  \label{R}
\end{equation}

The reflection amplitudes can be easily expressed through the two point
functions of boundary fields:

\begin{equation}
D(a)=\left\langle B_a(0),B_a(x)\right\rangle |x|^{2\Delta (a)}.  \label{D}
\end{equation}
For the evaluation of two point functions we can use (following the lines of
section 2) the screening charges which commute with all generators of $W$
-symmetry. In boundary NATT there are two types of screening charges 
$\widehat{Q}_H(i)$ and $\widehat{Q}_B(i)$:

\begin{equation}
\widehat{Q}_H(i)=\mu \int\limits_Hd^2xV_{be_i}(x,y);\quad \widehat{Q}
_B(i)=\mu _B\int dxB_{be_i}(x),  \label{S}
\end{equation}
where $\mu _B$ is given by eq.(\ref{MND}).

Using these screening charges we can express the structure constants of the
OPE of fields $B_a$ with strongly degenerate boundary fields in terms of
Coulomb gas integrals (see ref.\cite{FZZA} for details). The OPEs of fields 
$B_a$ with these fields contain only the finite number of primary fields. The
simplest strongly degenerate fields in NATT are the fields $B_{-b\omega _i}$,  
where $\omega _i$ are the fundamental weights of $G$. For the calculation
of boundary two point functions $D(a)$ it is convenient, following \cite{TEC}, 
\cite{FZZA}, to consider the auxiliary three point functions, including these
fields:

\begin{equation}
\left\langle B_a(x_1)B_{a+b\omega _i}(x_2)B_{-b\omega _i}(x)\right\rangle.
\label{AU}
\end{equation}
Then, tending $x\rightarrow x_2$ we can express the asymptotics in terms of
function $D(a)$ multiplied by the structure constant, which in the usual
normalization is equal to one. Instead, tending $x\rightarrow x_1$ we can
express the asymptotics in terms of function $D(a+b\omega _i)$ multiplied by
the structure constant ${\bf C}_{a,-b\omega _i}^{a+b\omega _i}$, which can
be calculated using the screening charges (\ref{S}). Equating these two
expressions we obtain:

\begin{equation}
\frac{D(a)}{D(a+b\omega _i)}={\bf C}_{a,-b\omega _i}^{a+b\omega _i}
\label{SC}
\end{equation}
where:

\begin{eqnarray}
{\bf C}_{a,-b\omega _i}^{a+b\omega _i}&=&
\left[ \frac{\pi \mu 2^{-2-2b^2}}{-\gamma (-b^2)}
\right] ^{\omega _i\cdot \rho }
\prod\limits_{\alpha>0}
\prod\limits_{k=1}^{\alpha \cdot \omega _i}
\frac{\Gamma ((b\overline{a}_\alpha +(k-1)b^2)/2)}
     {\Gamma ((1-b\overline{a}_\alpha -(k-1)b^2)/2)}
\nonumber \\
&& \times
\frac{\Gamma (-(b\overline{a}_\alpha +kb^2)/2)}
     {\Gamma ((1+b\overline{a}_\alpha +kb^2)/2)}
\label{C}
\end{eqnarray}
here $\bar{a}=(a-Q),\bar{a}_\alpha =(a-Q)\cdot \alpha $ like and in eq.(\ref
{abar}).

To construct a solution to these functional equations it is convenient to
use the special function ${\bf G}(x)$ (see for example \cite{FZZA}), which
is self dual entire function with zeroes at $x=-nb-m/b;n,m=0,1,2...$ and
enjoys the following shift relations:

\begin{equation}
{\bf G}(x+b)=\frac{b^{1/2-bx}}{\sqrt{2\pi }}\Gamma (bx){\bf G}(x);\quad
\quad {\bf G}(x+1/b)=\frac{b^{x/b-1/2}}{\sqrt{2\pi }}\Gamma (x/b){\bf G}(x).
\label{GAM}
\end{equation}
The integral representation which is valid for $\mathop{\rm Re}x>0$, has a form:

\begin{equation}
\log {\bf G}(x)=\int\limits_0^\infty \frac{dt}t\left[ \frac{e^{-qt/2}-e^{-xt}
}{(1-e^{-bt})(1-e^{t/b})}+\frac{(q/2-x)^2}2e^{-t}+\frac{q/2-x}t\right]
\label{IR}
\end{equation}
where:

\begin{equation}
q=b+1/b.  \label{Q}
\end{equation}
With this function the solution to the functional equations (\ref{SC},\ref{C}), 
satisfying the normalization condition $D(a)D(2Q-a)=1$ can be written in
the form:

\begin{equation}
D(a)=\frac{A_B(2Q-a)}{A_B(a)}  \label{DA}
\end{equation}
where:

\begin{eqnarray}
A_B(a)&=&\left( \pi \mu \gamma (b^2)b^{2-2b^2}\right) ^{\bar{a}\cdot\rho/2b}
\nonumber \\
&&\times
\prod\limits_{\alpha>0}\frac{{\bf G}(\bar{a}_\alpha ){\bf G}^2((q-
\bar{a}_\alpha )/2)}{{\bf G}((q+b+\bar{a}_\alpha )/2){\bf G}((q-b+\bar{a}
_\alpha )/2)}.  \label{AB}
\end{eqnarray}
Reflection amplitude for arbitrary element $\widehat{{\bf s}}$ of the Weyl
group of Lie algebra $G$ can be written as:

\begin{equation}
{\cal R}_s(a)=\frac{A_B(s(a))}{A_B(a)}=\frac{A_B(Q+\widehat{{\bf s}}(a-Q))}
{A_B(a)}.  \label{RS}
\end{equation}
These equations describe two point functions and reflection amplitudes for
NATT with boundary conditions (\ref{MND}). To obtain the same values for the
dual theory, which corresponds to the Neumann boundary conditions (\ref{N})
we should change in eq.(\ref{AB}) $b\rightarrow 1/b$ and transform the bulk
parameter $\mu \rightarrow \widetilde{\mu }$, where $\pi \mu \gamma (b^2)=$ 
$(\pi \widetilde{\mu }\gamma (1/b^2))^{b^2}$. We note that unlike boundary
reflection amplitudes the bulk reflection amplitudes (\ref{rsa},\ref{amp})
are invariant under this transformation.

Boundary reflection amplitudes (\ref{AB},\ref{RS}) can be used for the
calculation of vacuum expectation values of the boundary exponential fields
in ATTs, in the same way as it was done in section 3 
for ATTs defined at the whole
plane. Here we adopt the conventual normalization of boundary exponential
fields (see e.g. \cite{FLZ}) corresponding to the short distance asymptotics
at $\left| x_1-x_2\right| \rightarrow 0$:

\[
e^{a\varphi /2}(x_1)e^{a\varphi /2}(x_2)=\left| x_1-x_2\right| ^{a^2}+... 
\]
We define boundary one point function $G_B(a)$ as:

\begin{equation}
G_B(a)=\left\langle \exp (a\cdot \varphi /2)\right\rangle _B  \label{OP}
\end{equation}
For any element $\widehat{{\bf s}}$ of Weyl group ${\cal W}$ this function
satisfies the functional reflection relation: 
\begin{equation}
G_B(a)={\cal R}_s(a)G_B(s(a)).  \label{FE}
\end{equation}

The minimal meromorphic solution to these functional equations which 
respects all
the symmetries of extended Dynkin diagrams of Lie algebras $ADE$ has the
form:

\begin{eqnarray}
G_B(a)&=&
\left[ \frac{mk(G)\Gamma \left( \frac 1{(1+b^2)h}\right) \Gamma
\left( \frac{b^2}{(1+b^2)h}\right) b^2}{2\Gamma \left( \frac 1h\right)
h(1+b^2)}\right] ^{-a^2/2}
\nonumber \\
&& \times
\exp \left( \int\limits_0^\infty \frac{dt}t[\frac{a^2}2
e^{-2t}-F_B(a,t)]\right)  \label{vevs}
\end{eqnarray}
with

\begin{equation}
F_B(a,t)=f(t)\sum_{\alpha >0}\frac{\sinh (ba_\alpha t)\sinh ((b(a-2Q)_\alpha
+h(1+b^2))t)}{\sinh 2t\sinh (2b^2t)\sinh ((1+b^2)ht)}  \label{fbs}
\end{equation}
where for boundary conditions (\ref{MND}) function $f(t)$ is:

\begin{equation}
f(t)=2e^t\sinh ((1+b^2)t)\cosh (b^2t)  \label{fB}
\end{equation}
and for dual theory which corresponds to Neumann boundary conditions (\ref{N})
we should do the substitution $f(t)\rightarrow \widetilde{f}(t)$:

\begin{equation}
\widetilde{f}(t)=2e^{tb^2}\sinh ((1+b^2)t)\cosh t.  \label{fN}
\end{equation}

It is easy to see from the explicit form of $G_B(a)$ that in the classical
limit ($b\rightarrow 0,b\varphi $ is fixed) the boundary VEV 
$\widetilde{\varphi }_{0,B}$ of the field $b\varphi $ for the Neumann boundary
conditions coincides with 
classical vacuum $\widetilde{\varphi }_0$ (\ref{clv}) in the bulk. 
For the boundary conditions
(\ref{MND}) it happens only for Lie algebra $A_r$, where both these values
vanish. For other cases we can derive from eqs.(\ref{vevs}-\ref{fB}) that

\begin{equation}
\widetilde{\varphi }_{0,B}=\widetilde{\varphi }_0+\vartheta  \label{TET}
\end{equation}
where:

\begin{equation}
\vartheta =-\sum\limits_{\alpha >0}\alpha \int\limits_0^\infty \frac{dt}t
\frac{\sinh ((h-2\rho
_\alpha )t)}{\sinh (ht)}\tanh t.  \label{TT}
\end{equation}
These integrals can be calculated explicitly and expressed in terms of the
logarithms of the trigonometric functions of the parameter $\pi /h$. 
Vector $\vartheta $ is simply related with boundary soliton solution which
describes the classical vacuum configuration. The classical problem for this
solution $\phi (y),y>0$ can be formulated in the following way. We are
looking for the solution to classical Toda equations, which decrease at $
y\rightarrow \infty $ and satisfies at $y=0$ the boundary conditions that
follow from action (\ref{atb}). After rescaling and shifting (see section 4)
the field $\phi =b\varphi -$ $\widetilde{\varphi }_0$ satisfies the equation:

\begin{equation}
\partial _y^2\phi =m^2\sum_{i=0}^rn_ie_i\exp (e_i\cdot \phi );\qquad n_0=1
\label{EQT}
\end{equation}
and boundary condition at $y=0$:

\begin{equation}
\partial _y\phi =m\sum\limits_{i=0}^r\sqrt{n_i}e_i\exp (e_i\cdot \phi /2).
\label{YN}
\end{equation}
The vector $\vartheta $ is equal to the boundary value of this solution: 
$\vartheta =\phi (0),$ and hence, it completely fixes the solution.

We note that eq.(\ref{EQT}) has $r$ independent integrals. 
These integrals give the
equations to parameter $\vartheta $. In particular, it is easy to derive from
the first integral that numbers $E_i=\exp (e_i\cdot\vartheta /2),i=0,1...,r$ 
should satisfy the following ''sum rules'':

\begin{equation}
\sum\limits_{i=0}^r\sum\limits_{j=0}^r\left( 2\delta _{ij}-e_i\cdot e_j\right) 
\sqrt{n_i}\sqrt{n_j}E_iE_j=2h.  \label{SR}
\end{equation}
Numbers $E_i$ possess all the symmetries of extended Dynkin diagram of Lie
algebra $G$. Together with these symmetries the ''sum rules'' fix
completely parameters $E_i$ (and hence vector $\vartheta $) for Lie algebras 
$D_4$ and $D_5$. Consider, for example, the second case. There we have that 
$E_0=E_1=E_4=E_5=u,$ and $E_2=E_3=v$. But numbers $E_i$ are not independent.
They satisfy the condition: $E_0^{-1}=E_1E_4E_5E_2^2E_3^2,$ or $u=1/v.$ Then
we find from eq.(\ref{SR}) that $4\sqrt{2}+2v^2=8$. This defines all
parameters $E_i$. The result is in perfect agreement with eq.(\ref{TT}),
which for Lie algebra $D_r$ can be rewritten as:

\[
E_0^2=E_1^2=E_{r-1}^2=E_r^2=\frac 8{h^2\sin ^2(\pi /h)}; 
\]

\begin{equation}
E_k^2=\exp \left( \int\limits_0^\infty \frac{dx}x\frac{4\sinh ^2x\cosh 2(r-2k)x}{\sinh
2(r-1)x\cosh 2x}\right) ,\quad k=2,...,r-2.  \label{ESQW}
\end{equation}
For other Lie algebras we can only check that vector $\vartheta $ defined by
eq.(\ref{TT}) satisfies eq. (\ref{SR}).

The solution to the eqs. (\ref{EQT}),(\ref{YN}) can be expressed in terms
of tau-functions associated with multi-soliton solutions of classical ATT
equations (see, for example \cite{BOWC},\cite{BOW}). For $D$ and $E$ series
of algebras (besides the cases $D_4$ and $D_5$ \cite{BOWC} ) the explicit
form of these solutions is not known. We suppose to discuss these solutions
in more details in the separate publication. Here we consider the classical
boundary ground state energy which can be defined as:

\begin{equation}
\mathcal{E}_{bound}^{(cl)}=\frac 1{4\pi b^2}[2m\sum_{i=0}^r\sqrt{n_i}%
E_i+\int\limits_0^\infty dy(\frac 12(\partial _y\phi
)^2+m^2\sum_{i=0}^r n_i(e^{e_i\cdot \phi }-1))].  \label{eBon}
\end{equation}
We note that numerical values of $\vartheta ^2$, defined by eq.(\ref{TT})
are rather small for all $G$, and the integral term in eq.(\ref{eBon}) can
be calculated with the good accuracy using the bilinear approximation:

\[
\int\limits_0^\infty dy(\frac 12(\partial _y\phi
)^2+m^2\sum_{i=0}^r n_i(e^{e_i\cdot \phi }-1))=\frac
m2\sum_{a,b}(M^{1/2})_{ab}\vartheta _a\vartheta _b+O(|\vartheta |^3) 
\]
where $M_{ab}$ is the mass matrix of ATT, defined by eq.(\ref{mab}).

More careful analysis of the eqs.(\ref{EQT}),(\ref{YN}) gives us the
reasons to write the expression for boundary ground state energy in the
following form. Namely, we denote as $\Sigma _m(G)$ the sum of masses of all
particles in ATT:

\begin{equation}
\Sigma _m(G)=\sum_{i=1}^rm_i=m\sum_{i=1}^r\nu _i=m\cdot tr(M^{1/2}).
\label{sIg}
\end{equation}
It has the following values for simply-laced Lie algebras:

\begin{eqnarray}
\Sigma _m(A_{n}) &=&2m\cot (\pi /2h);\quad 
\Sigma _m(D_n)=\frac{2m\cos
(\pi /4-\pi /2h)}{\sin (\pi /2h)};  \nonumber \\
\Sigma _m(E_6) &=&m(6-2\sqrt{3})^{1/2}
\frac{\cos (\pi /8)}{\sin (\pi /2h)};
\quad \Sigma _m(E_7)=\frac{2m\sin (2\pi /9)}{\sin (\pi /2h)}; 
 \nonumber \\
\Sigma _m(E_8) &=&4m(\sqrt{3}\sin (\pi /5)\sin (\pi /30))^{1/2}\frac{\cos
(\pi /5)}{\sin \left( \pi /2h\right) }.  \label{sUm}
\end{eqnarray}
The classical boundary ground state energy (\ref{eBon}) can be written in
terms of these values as:

\begin{equation}
\mathcal{E}_{bound}^{(cl)}(G)=\frac h{4\pi b^2}\tan (\pi /2h)\Sigma _m(G).
\label{eCl}
\end{equation}

In quantum case the boundary ground state energy $\mathcal{E}_{bound}^{(q)}$
will have the contributions coming from the boundary term in the Hamiltonian
and from the bulk fluctuations around the background solution. The
contributions of the first type can be calculated using the explicit
expression for vacuum expectation values (\ref{vevs}). For small $b$ the
first quantum correction of the second type can be expressed through the
boundary $S-$matrix at $b=0$ (see, for example \cite{COD},\cite{COT}). We
will discuss these boundary $S$-matrices bellow. Here we note that at the
strong coupling region $b>>1$ our theory is described by the weakly coupled
dual ATT with Neumann boundary conditions (\ref{N}). At strong coupling
limit the dual theory is the set of $r$ free bosonic theories with masses $m_i$.
The boundary ground state energy for free massive bosonic theory with
Neumann boundary conditions and mass $m_i$ can be easily calculated and is
equal to $m_i/8$. The first perturbative correction in the weakly coupled
dual theory can be also evaluated with the result:

\begin{equation}
\mathcal{E}_{bound}^{(q)}(G)=\frac{\Sigma _m(G)}8\left( 1+\frac \pi
{2hb^2}\cot (\pi /2h)+O(1/b^4)\right) .  \label{eDu}
\end{equation}
Both asymptotics $b\rightarrow 0$ (\ref{eCl}) and $b\rightarrow \infty $ (%
\ref{eDu}) are in agreement with the following conjecture for boundary
ground state energy:

\begin{equation}
\mathcal{E}_{bound}^{(q)}(G)=\frac{\sin (\pi /2h)\Sigma _m(G)}
{8\sin (\pi x/2h)\cos (\pi (1-x)/2h)}  \label{eQ}
\end{equation}
where:

\begin{equation}
x=\frac{b^2}{1+b^2}=\frac{B}{2}=\frac bq.  \label{xB}
\end{equation}

The nonperturbative check of this conjecture can be done using the boundary
Thermodynamic Bethe Ansatz equations \cite{SSM}. The kernels in these
nonlinear integral equations depend on the bulk and boundary $S$-matrices.
The boundary $S$-matrix (reflection coefficient) in ATT for the particle $j$
corresponding to the fundamental representation $\pi _j(G)$ can be defined
as:

\begin{equation}
|j,-\theta \rangle _{out}=K_j(\theta )|j,\theta \rangle _{in}  \label{kI}
\end{equation}
where $\theta $ is the rapidity of particle $j$.

The reflection coefficients for $A_{n-1}$ ATT with boundary conditions
discussed above were conjectured in \cite{CODS},\cite{DELG}. They can be
written in terms of function:

\begin{equation}
\left( z\right) =\frac{\sin \left( \frac \theta {2i}+\frac \pi {2h}z\right) 
}{\sin \left( \frac \theta {2i}-\frac \pi {2h}z\right) }  \label{Z}
\end{equation}
in the following way:

\begin{equation}
K_j(\theta )=\prod\limits_{a=1}^j(a-1)(a-n)(-a+x)(-a-n+1-x).  \label{aNk}
\end{equation}
Unfortunately, we were not able to find in the literature the conjecture for
other Lie algebras consistent with duality properties
discussed above. 
So, we will give here the conjecture which naturally
generalizes the reflection coefficients (\ref{aNk}) to the other
simply-laced Lie algebras. To do this we rewrite eq. (\ref{aNk}) in the form:

\begin{equation}
K_j(\theta )=\exp (-i\delta _j(\theta ))  \label{kJ}
\end{equation}
where

\begin{equation}
\delta _j=\int\limits_0^\infty \frac{dt}t\sinh (\frac{2h\theta t}\pi )
[\sinh ((1-x)t)\sinh((h+x)t)
\Delta _j(A_{n-1},t)-2]  \label{dJ}
\end{equation}
and

\begin{equation}
\Delta _j(A_{n-1},t)=\frac{8\sinh \left( jt\right) \sinh ((n-j)t)}
{\sinh(t)\sinh (2ht)}.  \label{DJ}
\end{equation}
The natural generalization of these equations can be written as:

\begin{equation}
K_j(\theta )=\Phi _j(\theta )\exp (-i\delta _j(\theta ))  \label{gNr}
\end{equation}
where $\Phi _j(\theta )$ are CDD factors, satisfying the conditions:

\begin{equation}
\Phi _j(\theta )\Phi _j(-\theta )=1;\quad \Phi (\theta )
\Phi (\theta +i\pi)=1  \label{cDd}
\end{equation}
and function $\delta _j\left( \theta \right) $ 
is defined by eq.(\ref{dJ})
with the substitution 
$\Delta _j(A_{n-1})\rightarrow $ $\Delta _j(G,t)$,
where:

\begin{equation}
\Delta _j(G,t)=\frac 4{\cosh \left( ht\right) }\left( (2\cosh (t)-2)
\delta_{mn}+e_m\cdot e_n\right) _{jj}^{-1}.  \label{cOn}
\end{equation}
The most important part of this conjecture is that for the particles $j$
corresponding to fundamental representations $\pi _j(G)$ with the
fundamental weights satisfying the condition $-e_0\cdot \omega _j=1$ 
(or $ n_j $ in eq.(\ref{mr}) is equal to $1$), 
the CDD factors $\Phi _j(\theta )$
in eq.(\ref{gNr}) are equal to one.

This statement fixes completely the boundary $S$-matrix for Lie 
algebras $D_n,E_6,E_7$. We denote as $K_f(G,\theta )$ 
the fundamental reflection
coefficients. It means that all other amplitudes $K_j(\theta )$ 
can be obtained from fundamental reflection factors
$K_f(G,\theta )$ by the application 
of boundary bootstrap fusion procedure 
\cite{GZ},\cite{FK}. It is easy to see that 
$K_f(D_n,\theta )=K_n(\theta)=K_{n-1}(\theta )$, 
where particles $n$ and $n-1$ correspond to the spinor
representations of $D_n$; $K_f(E_6,\theta )=K_1\left( \theta \right) =
K_{\overline{1}}\left( \theta \right) $, 
where particles $1$ and $\overline{1}$
form the doublet of lightest particles in $E_6$ ATT and $K_f(E_7,\theta
)=K_1(\theta )$, where $1$ is the lightest particle in $E_7$ ATT. For all
these three cases the CDD factors $\Phi _f(\theta )=1$ and the reflection
coefficients are given by eqs.(\ref{kJ})(\ref{dJ}) with $j=f$ and

\begin{equation}
\Delta _f(G,t)=\frac{8\sinh \left( ht/2\right) \sinh ((h/2+q-1)t)}{\sinh
(qt)\sinh (2ht)}  \label{dE}
\end{equation}
where $q(G)=\max_in_i;q(D)=2,q(E_6)=3,q(E_7)=4$.

As an example of the application of boundary bootstrap equations we give
here the reflection coefficients $K_j$ for the particles 
$j=1,2,...,n-2$ in $ D_n$ ATT, which can be obtained 
from the amplitude $K_f\left( \theta
\right) =K_n(\theta )=K_{n-1}(\theta )$. These functions can be written in
the form (\ref{gNr}), where:

\[
\Delta _j(D_n,t)=\frac{16\sinh jt\cosh ((h/2-j)t)
\sinh (ht/2)}{\sinh t\sinh2ht}
\]
and CDD factor 
\[
\Phi _j=\exp \left( i\int\limits_0^\infty dt\frac{8\sinh (2h\theta t/\pi )\sinh
((1-x)t)\cosh xt\sinh ((j-1)t)\sinh jt}{t\sinh 2t\cosh ht}\right) 
\]

Lie algebra $E_8$ has no fundamental representations with $n_j=1$. The
lightest particle in this case is associated with adjoint representation and 
$K_f(E_8,\theta )=K_{ad}(\theta )$. The adjoint representations for Lie
algebras $D,E$ have $n_{ad}=2$ and, hence, CDD factors should appear. The
reflection coefficients $K_{ad}(\theta )$ for Lie algebras $D$ and $E$ can be 
written in the the form (\ref{gNr}), with:

\[
\Delta _{ad}(G,t)=
\frac{8\cosh ((q-1)t)\cosh ((h/2-q)t)}{\cosh (ht/2)\cosh ht}
\]
where $q(G)$ is defined above ($q(E_8)=6$) and CDD factor is:

\[
\Phi _{ad}(\theta )=\exp \left( i\int\limits_0^\infty dt
\frac{8\sinh (2h\theta t/\pi )\sinh((1-x)t)\cosh xt\sinh t}
{t\cosh ht}\right) .
\]

The analysis of the boundary Thermodynamic Bethe Ansatz equations with
kernels depending on the reflection coefficients written above gives the
exact agreement with eq.(\ref{eQ}) for the boundary ground state energy.

At the end of this section we note that semiclassical limit of function $
G_B(a)$ contains the information about the short distance asymptotics for
the following classical boundary problem. Let $\phi (x,y)$ be a smooth
function that for $y>0$ solves the equations:

\begin{displaymath}
(\partial _y^2+\partial _x^2)\phi =m^2\sum_{i=0}^rn_ie_i\exp (e_i\cdot \phi
);\quad \partial _y\phi (x,0)=m\sum\limits_{i=0}^r\sqrt{n_i}e_i\exp
(e_i\cdot \phi /2)  
\end{displaymath}
and satisfies the following asymptotic conditions:

\begin{equation}
\phi \rightarrow 0,\quad x^2+y^2\rightarrow \infty ;\quad \phi \rightarrow
-a\log (x^2+y^2)+{\cal B}(a),\quad x^2+y^2\rightarrow 0.  \label{ACN}
\end{equation}
Then, exactly at the same way as it was done in section 4, we can derive
that ${\cal B}(a)=C(a)-\widetilde{\varphi }_0(0),$ where:

\begin{equation}
C(a)=\lim_{b\rightarrow 0}2b^2\partial _a\log G_B(a/b).  \label{BAAS}
\end{equation}
For Neumann boundary conditions (\ref{fN}) we obtain that ${\cal B}(a)=B(a)$, 
where $B(a)$ is given by eq.(\ref{baf}). It is in agreement with a fact that
cylindrical solutions, studied in section 4, satisfy the Neumann boundary
conditions.

\section{Integrable Deformations of Toda Theories and Duality}

The duality plays an important role in the analysis of statistical and
quantum field theory (QFT) systems. It maps a weak coupling region of one
theory to a strong coupling region of the other and makes it possible to use
perturbative and semiclassical methods for the study of dual systems in
different regions of the coupling constants. For example, a well known
duality between sine-Gordon and massive Thirring model \cite{CM} together
with integrability plays a crucial role for the justification of the exact
S-matrix for these QFT . Another interesting example of the duality in two
dimensional integrable systems is the weak coupling-strong coupling flow
from the affine Toda theories to the same theories with the dual
affine Lie algebra \cite{CD}. The example of duality in the boundary Toda
theories was considered in section 5. The phenomenon of electric-magnetic
duality in four dimensional gauge theories conjectured in \cite{MNO} and
developed in \cite{SW} opens the possibility for the nonperturbative
analysis of the spectrum and the phase structure in the supersymmetric
Yang-Mills theories.

Known for many years the phenomenon of duality in QFT still looks rather
mysterious and needs further analysis. This analysis essentially simplifies
for the 2-d integrable relativistic theories. These QFTs besides the
Lagrangian formulation possess also the unambiguous definition in terms of
factorized scattering theories (FST). The FST, i.e. the explicit description
of the spectrum of particles and their scattering amplitudes, contains all
the information about the QFT. These data permit one to use nonperturbative
methods for the calculation of the observables in the integrable theories. The
comparison of the observables calculated from the FST data and from
perturbative or semiclassical analysis based on the Lagrangian formulation
makes it possible in some cases to justify the existence of two different
(dual) representations for the Lagrangian description of the theory.

In the previous sections we considered simply laced Toda theories which are
self-dual. In this section we briefly discuss the massive and conformal
field theories, which can be considered as integrable deformations of affine
and non-affine Toda theories. We describe duality properties of these
theories and calculate the reflection amplitudes for the conformal case.
These QFTs for massive case where introduced and studied in (\cite{FA}).
They form three series of QFTs, numerated by index $\sigma =1,2,3$ (we
reserve $\sigma =0$ for unperturbed CFTs). These QFTs can be described by the
scalar field $\Phi $ and $r$-component field $\varphi =\left( \varphi
_{1,...,}\varphi _r\right) $ with the action:

\begin{equation}
{\cal A}_r^{(\sigma )}=\int d^2x \left [ \frac{(\partial _\mu \Phi )^2}{8\pi }+
\frac{(\partial _\mu \varphi )^2}{8\pi }+2\mu ^{\prime }\cos (\gamma \Phi
)e^{be_r\cdot \varphi } 
 +\mu \sum\limits_{i=1}^{r-1}e^{be_i\cdot \varphi}
+U_\sigma (\varphi ) \right]  \label{dTT}
\end{equation}
where $e_i$ are the simple roots $(e_i\cdot \varphi =\varphi _i-\varphi
_{i+1},i\leq r-1;e_r\cdot \varphi =\varphi _r)$ of Lie algebra $B_r$ and the
parameters $b$ and $\gamma $ satisfy the relation:

\begin{equation}
\gamma ^2-b^2=1.  \label{aG}
\end{equation}
For these values of parameters the QFT (\ref{dTT}) is integrable for three
different perturbations $U_\sigma $ of the CFT, corresponding to $U_\sigma
=0.$ Namely:

\begin{equation}
U_1(\varphi )=\mu _1e^{-2b\varphi _1};\quad U_2(\varphi )=\mu _2e^{-b\varphi
_1}\quad U_3(\varphi )=\mu _3e^{-b(\varphi _1+\varphi _2)}.  \label{Us}
\end{equation}
The integrability of these QFTs, was proved in \cite{FA} by explicit
construction of nontrivial quantum integrals. It is interesting to note that
corresponding classical theory is not integrable. The reason is that due to
condition (\ref{aG}) the coupling constant $\gamma $ is not small and we can
not reach the classical limit. To understand this limit and to do the QFTs  
(\ref{dTT}) suitable for the prturbative analysis it is convenient to use 2D
fermion-boson correspondence \cite{CM} and to rewrite the action (\ref{dTT})
in the form of massive Thirring model coupled with ATT:

\begin{eqnarray}
{\cal A}_r^{(\sigma )}&=&\int d^2x[\overline{\psi }i\gamma _\mu \partial
_\mu \psi -\frac{b^2}{2(1+b^2)}(\overline{\psi }\gamma _\mu \psi )^2+\pi \mu
^{\prime }\overline{\psi }\psi e^{be_r\cdot \varphi }
\nonumber \\
&&
+\frac 1{8\pi }(\partial _\mu \varphi )^2+\frac{\pi \mu ^{\prime 2}}{4b^2}
e^{2be_r\cdot \varphi }+\mu \sum\limits_{i=1}^{r-1}e^{be_i\cdot \varphi
}+U_\sigma (\varphi )]. \label{acT} 
\end{eqnarray}
We added to the action the exponential term $\pi \mu ^{\prime
2}/4b^2e^{2be_r\cdot \varphi }$ as the usual contact contact counterterm to
cancel the divergencies coming from fermion loop, however, this term becomes
important in the weak coupling (semiclassical) limit. Near the classical 
vacuum of the QFT (\ref{acT})  the parameter $\mu \sim \mu
^{\prime 2}/b^2$ (the same is true for parameters $\mu _\sigma $ in 
$U_\sigma $) and we can neglect the fermionic terms in the action 
which do not contain the
derivatives. The first term with derivatives can be again bosonized, but it
completely decouples in this limit. The classical part of the action is now
described by non-simply lased ATT which is, of course, integrable. The
corresponding background semiclassical CFT ($U_\sigma =0$) is described by
NATT with Lie algebra $C_r$.

Integrability imposes strong limitation to the scattering amplitudes (Yang
Baxter equations) and permits us to fix completely $S$-matrix of the 
QFTs (\ref{dTT},\ref{acT}). 
Perturbative calculations, analysis of the FST and the Bethe
ansatz technique were used in (\cite{FA}) to show that these field theories
possess the dual representation available for the perturbative analysis in
the strong coupling limit, when $\widetilde{b}=1/b\rightarrow 0$. The dual
theory can be formulated as the nonlinear sigma-model with Witten's
Euclidean black hole metric coupled with non-simply laced ATT. Lie algebras
associated with these ''dual'' Toda theories belong to the dual series of
affine algebras but have a smaller rank equal to $r-1$.

To describe the action of the dual theory we introduce the complex scalar
field $\chi =\chi _1+i\chi _2$ and Toda field $\phi $ with $r-1$ components: 
$\phi =\left( \phi _1,...,\phi _{r-1}\right) $. In terms of these fields the
dual action has the form:

\begin{equation}
\widetilde{{\cal A}}_r^{\left( \sigma \right) }=\int d^2x\left[ \frac 1{2\pi 
\widetilde{b}^2}\frac{\partial _\mu \chi \partial _\mu \overline{\chi }}
{1+\chi \overline{\chi }}+\frac{(\partial _\mu \phi )^2}{8\pi }+2\widetilde
{\mu }\chi \overline{\chi }e^{\widetilde{b}e_{r-1}\cdot\phi }
+\widetilde{\mu }
\sum\limits_{i=1}^{r-1}e^{\widetilde{b}e_i\cdot\phi }+
\widetilde{U}_\sigma \right]
\label{Adu}
\end{equation}
where $e_i$ are the simple roots of Lie algebra $B_{r-1}$, $\widetilde{b}
=1/b $, and dual integrable perturbations are:

\begin{equation}
\widetilde{U}_1(\phi )=\widetilde{\mu }_1e^{-\widetilde{b}\phi _1};\quad 
\widetilde{U}_2(\phi )=\widetilde{\mu }_2e^{-2\widetilde{b}\phi _1};\quad 
\widetilde{U}_2(\phi )=\widetilde{\mu }_3e^{-\widetilde{b}(\phi _1+\phi _2)}.
\label{Udu}
\end{equation}

We see that the charged particles in these QFTs being weakly coupled
fermions (at small $b$) flowing to the strong coupling region (small 
$\widetilde{b}$ ) take one degree of freedom from Toda lattice and transform
to the weakly coupled bosons. This property of integrable QFTs (\ref{acT},\ref
{Adu}) is used in condensed matter physics for nonperturbative analysis of
superconductors coupled to phonons living in an insulating layer (see e.g. 
\cite{TC}). For nonperturbative analysis of these QFTs we need besides the
FST data, which were described in (\cite{FA}), also the CFT data,
characterizing the background CFTs.

The corresponding CFTs are described by the actions ${\cal A}_r^{(0)}$ eq.(\ref
{dTT}) and $\widetilde{{\cal A}}_r^{\left( 0\right) }$ eq.(\ref{Adu}) with 
$U_\sigma =$ $\widetilde{U}_\sigma =0$. From the duality of perturbed
theories it follows that these CFTs are also dual i.e. describe the same
theory. The CFT (\ref{Adu}) for $r=1$ was used in papers \cite{W},\cite{DVV}
for the description of the string propagating in a black hole background.
For arbitrary $r$ these CFTs, known as non-abelian Toda theories, were
considered in \cite{GS} as the models for extended black holes. In many
cases, however, the dual representation with action ${\cal A}_r^{(0)}$,
which can be called Sine-Toda theory is more convenient for the analysis.

The conformal invariance of Sine-Toda theories is generated by holomorphic
stress-energy tensor:

\begin{equation}
T(z)=-\frac 12\left( \partial _z\Phi \right) ^2-\frac 12(\partial _z\varphi
)^2+Q\partial _z^2\varphi  \label{SEt}
\end{equation}
with $Q=\rho /b+b\rho ^{\prime }$, where $\rho $ and $\rho ^{\prime }$ are
the Weyl vectors of Lie algebras $B_r$ and $D_r$ respectively. The
exponential fields (for simplicity we consider the spinless fields):

\begin{equation}
V_{a,\eta }(x)=\exp (a\cdot \varphi +i\eta \Phi )  \label{Vah}
\end{equation}
are conformal primary field with dimensions:

\begin{equation}
\Delta (a,\eta )=\frac 12(\eta ^2+a\left( Q-a\right) ). \label{Dimen} 
\end{equation}
In particular, fields $V_{be_i,0};i=1,...,r-1,V_{be_r,\pm \gamma }$ have
conformal dimensions equal to one.

Besides the conformal symmetry, generated by $T\equiv T_2$ these CFTs
possess an infinite-dimensional symmetry generated by the chiral algebra 
${\cal T}$, which includes an infinite number of holomorphic fields $T_j$
with integer spins. The detailed description of the chiral algebra ${\cal T}$
is not in the scope of this paper. As an example, we give here the spin-3
field $T_3\in {\cal T}$ for the case $r=1$. In this case the theory is
described by two fields $\Phi $ and $\varphi _1$ and can be called as
Sine-Liouville CFT with an action:

\begin{equation}
{\cal A}_1^{(0)}=\int d^2x\left[ \frac{(\partial _\mu \Phi )^2}{8\pi }+\frac{
(\partial _\mu \varphi _1)^2}{8\pi }+2\mu ^{\prime }\cos (\gamma \Phi
)e^{b\cdot \varphi _1}\right].  \label{SLm}
\end{equation}
The holomorphic field $T_3$ for this CFT has a form:

\begin{eqnarray*}
T_3 &=&\frac{1+3b^2}{1+b^2}(\partial _z\Phi )^3+\frac{3b^2}2(\partial _z\Phi
)(\partial _z\varphi _1)^2+\frac{3b^3}2(\partial _z\Phi )(\partial
_z^2\varphi _1) \\
&&-\frac 32b(1+b^2)(\partial _z^2\Phi )(\partial _z\varphi _1)+\frac 14
(1+b^2)(\partial _z^3\Phi ).
\end{eqnarray*}
The other fields $T_j,j>3$ for this theory can be obtained by fusion of the
field $T_3.$

The fields $T_j$ in general case can be represented as the differential
polynomials of the fields $\partial _z\Phi ,\partial _z\varphi $ of weight 
$ j $. It means that the exponential fields (\ref{Vah}) are the primary fields
of chiral algebra ${\cal T}$. The corresponding eigenvalues $t_j(a,\eta )$
of the operators $T_{j,0}$ (zero Fourier components of currents $T_j$)
possess reflection symmetry: $t_j(a,\eta )=t_j(s(a),\eta )$ with respect to
action of Weyl group ${\cal W}$ : $s(a)=Q+\widehat{{\bf s}}(a-Q)$ of Lie
algebra $B_r.$ The fields $V_{a,\eta }$ and $V_{s(a),\eta }$ are then
related by reflection amplitudes:

\begin{equation}
V_{a,\eta }=R_s(a,\eta )V_{s(a),\eta }.  \label{Rre}
\end{equation}
For calculation of reflection amplitudes in Sine-Toda theory it is
convenient to use the screening charges:

\begin{displaymath}
\widehat{Q}_i=\int d^2x\exp (be_i\cdot \varphi ),i=1,...,r-1;\quad \widehat{Q
}_{\pm }=\int d^2x\exp (be_r\cdot \varphi \pm i\gamma \Phi )  
\end{displaymath}
which commute with all generators of chiral algebra. The calculation of the
reflection amplitudes $R_s(a,\eta )$ follows the lines of sections 2 and 5.
Here we give the result. To describe it is convenient to denote as $S$ $(L)$
the set of the short (long) positive roots of Lie algebra $B_r$: $(S:\alpha
>0,\alpha ^2=1),(L:\alpha >0,\alpha ^2=2)$. Then, for arbitrary reflection, 
$s(a)$ the reflection amplitude can be represented in the usual form:

\begin{equation}
R_s(a,\eta )=\frac{A_{s(a),\eta }}{A_{a,\eta }},  \label{Refl}
\end{equation}
and function $A_{a,\eta }$ is:

\begin{eqnarray}
A_{a,\eta } &=&\left( \frac{\pi \mu ^{\prime }}{2b^2}\right) ^{2\overline{a}
\cdot \omega _r/b}\ \left( \pi \mu \gamma (b^2)\right) ^{\overline{a}\cdot
\rho ^{\prime }/b}
 \prod\limits_{\alpha \in L}\Gamma (1-\overline{a}_\alpha b)\Gamma
(1-\overline{a}_\alpha /b) 
 \nonumber \\
&& \times
\prod\limits_{\alpha \in S}\frac{\Gamma (1-2\overline{a}
_\alpha b)\Gamma (1-\overline{a}_\alpha /b)}{\Gamma (1/2-\overline{a}_\alpha
b+\gamma \eta )\Gamma (1/2-\overline{a}_\alpha b-\gamma \eta )} 
\label{Aahg}
\end{eqnarray}
where $\omega _r$ is the fundamental weight of $B_r:2\omega _r\cdot
e_i=\delta _{r,i};\rho ^{\prime }$ is the Weyl vector of Lie algebra $D_r$
and $\overline{a}_\alpha =(a-Q)\cdot \alpha $.

It was noted in the beginning of this section that in the semiclassical limit 
$b\to 0$ the Sine-Toda CFT is effectively described by $C_r$ NATT and decoupled free field $\Phi$. In agreement with this in the limit $b\to 0$ with 
$\overline{a}/b$ fixed the reflection amplitudes (\ref{Refl}),(\ref{Aahg}) do 
not depend on parameter $\eta$ and coincide with reflection amplitudes for 
$C_r$ NATT calculated in \cite{BFAK}. 

Two points functions $D_r(a,\eta )$ of the fields $V_{a,\eta }$, normalized
by the condition $D_r(a,\eta )D_r(2Q-a,\eta )=1$ can be written as:

\begin{equation}
D_r(a,\eta )=\left| x\right| ^{4\Delta }\left\langle V_{a,\eta
}(x),V_{a,-\eta }(0)\right\rangle =\frac{A_{2Q-a,\eta }}{A_{a,\eta }}.
\label{Tpfr}
\end{equation}

Until now we considered spinless fields and non-compactified field $\Phi $.
For string theory applications it is, however, important to use the
periodicity property of Sine-Toda theories and to compactify the field $\Phi 
$ at the circle of length $2\pi /\gamma $, i.e. $\Phi =\Phi +2\pi k/\gamma
,k\in Z$. In this case the parameter $\eta $ is quantized: $\eta _n=\gamma n$. 
It is useful to introduce the ''dual'' field $\Phi ^{\prime }$, defined by
the relation: $\partial _\mu \Phi =\varepsilon _{\mu \nu}\partial_\nu \Phi
^{\prime }$ and to consider the local exponential fields with spin 
$\sigma =nm$, $m\in Z$, which can be written as:

\begin{equation}
V(a,\eta _{n,}\eta _m^{\prime })=\exp (a\cdot \varphi )\exp (i\eta _n\Phi
+i\eta _m^{\prime }\Phi ^{\prime })  \label{Sfiel}
\end{equation}
where $\eta _n=\gamma n$ and $\eta _k^{\prime }=m/2\gamma $. In particular,
field $\chi $ in non-Abelian Toda theories (\ref{Adu}) can be represented in
terms of the fields of Sine-Toda theories as:

\begin{equation}
\chi \sim \exp (-e_r\cdot \varphi /2b)\exp (i\Phi ^{\prime }/2\gamma ).
\label{Qhi}
\end{equation}
In the string theories associated with these CFTs the numbers $n$ and $m$
correspond to the winding number and momentum of sting propagating in the
black hole background. The total momentum is conserved. The total winding 
number is not conserved and the sum of winding numbers of the operators in 
$p$-point functions can take the values between $2-p$ and $p-2$. 

To obtain the reflection amplitudes and two point
functions for the fields $V(a,\eta _{n,}\eta _m^{\prime })$ we should do the
following substitution in the eq.(\ref{Aahg}). Namely,  
the dependence on parameter 
$\eta $ there appears in two $\Gamma $-function which are in the denominator
of the product over short roots. We should substitute $\eta \rightarrow \eta
_n+\eta _m^{\prime}$ in the argument of the first $\Gamma $-function and $\eta
\rightarrow \eta _n-\eta _m^{\prime}$ in the argument of the second one.  
For example,
for the Sine-Liouville theory ($r=1$ and $\overline{a}=\overline{a}_1$) we
obtain:

\begin{eqnarray}
D_1(a,\eta _n,\eta _m^{\prime}) &=&\left( \frac{\pi \mu ^{\prime }}{2b^2}\right)^{-2\overline{a}/b}\frac{\Gamma (1+2\overline{a}b)\Gamma (1+\overline{a}/b)}
{\Gamma (1-2\overline{a}b)\Gamma (1-\overline{a}/b)}\times  \label{TPfn} \\
&&\frac{\Gamma (1/2-\overline{a}b+\gamma (\eta _n+\eta _m^{\prime}))
\Gamma (1/2-\overline{a}b-\gamma (\eta _n-\eta _m^{\prime}))}
{\Gamma (1/2+\overline{a}b+\gamma
(\eta _n+\eta _m^{\prime}))
\Gamma (1/2+\overline{a}b-\gamma (\eta _n-\eta _m^{\prime})}. 
\nonumber
\end{eqnarray}
This two point function as well as duality between Sine-Liouville and
Witten's 2D black hole models were obtained in collaboration with
A.Zamolodchikov and Al.Zamolodchikov.

We note that besides the string theory, where two point functions of the
vertex operators $V(a,\eta _{n,}\eta _m^{\prime })$ contain the information
about the spectrum \cite{DVV} and partition function of the theory \cite{MOG}, 
reflection amplitudes,
derived in this section, can be used for the calculation of one point
functions and UV asymptotics in massive QFTs (\ref{dTT},\ref{acT},\ref{Adu}).
We suppose to discuss these problems in the separate publications.

\begin{center}
{\bf Acknowledgement}
\end{center}
I am very grateful to S.Lukyanov, A.Zamolodchikov and Al.
Zamolodchikov with whom many of similar results for another integrable QFT
were obtained. This work supported by part by the EU under contract ERBFRMX
CT 960012 and grant INTAS-OPEN-97-1312.

\end{document}